\documentclass[12pt,twocolumn]{emulateapj}

\usepackage{color}
\usepackage[usenames,dvipsnames]{xcolor}
\usepackage[colorlinks,urlcolor=blue,citecolor=blue,linkcolor=blue]{hyperref}
\usepackage{amsmath}
\slugcomment{}

\begin{document}

\title{Mapping the interstellar medium with near-infrared diffuse interstellar bands}
\shorttitle{Mapping the ISM with near-IR diffuse interstellar bands}

\author{G.~Zasowski\altaffilmark{1,2}, 
B.~M\'enard\altaffilmark{2,3,4}, \\ 
D.~Bizyaev\altaffilmark{5,6}, 
D.~A.~Garc\'{i}a~Hern\'{a}ndez\altaffilmark{7,8},
A.~E.~Garc\'{i}a~P\'{e}rez\altaffilmark{9},
M.~R.~Hayden\altaffilmark{6},
J.~Holtzman\altaffilmark{6}, \\
J.~A.~Johnson\altaffilmark{10}, 
K.~Kinemuchi\altaffilmark{5},
S.~R.~Majewski\altaffilmark{9},
D.~L.~Nidever\altaffilmark{11},
M.~Shetrone\altaffilmark{12}, 
J.~C.~Wilson\altaffilmark{9}
}
\shortauthors{Zasowski et al.}

\altaffiltext{1}{NSF Astronomy and Astrophysics Postdoctoral Fellow; gail.zasowski@gmail.com}
\altaffiltext{2}{Department of Physics \& Astronomy, Johns Hopkins University, Baltimore, MD, 21218, USA}
\altaffiltext{3}{Alfred P. Sloan Fellow}
\altaffiltext{4}{Kavli Institute for the Physics and Mathematics of the Universe, University of Tokyo, Kashiwa 277-8583, Japan}
\altaffiltext{5}{Apache Point Observatory, Sunspot, NM 88349, USA}
\altaffiltext{6}{Department of Astronomy, New Mexico State University, Las Cruces, NM 88003, USA}
\altaffiltext{7}{Instituto de Astrof\'{i}sica de Canarias, E-38205 La Laguna, Tenerife, Spain}
\altaffiltext{8}{Departamento de Astrof\'{i}sica, Universidad de La Laguna, E-38206 La Laguna, Tenerife, Spain}
\altaffiltext{9}{Department of Astronomy, University of Virginia, Charlottesville, VA 22904, USA}
\altaffiltext{10}{Department of Astronomy, The Ohio State University, Columbus, OH 43210, USA}
\altaffiltext{11}{Department of Astronomy, University of Michigan, Ann Arbor, MI 48104, USA}
\altaffiltext{12}{The University of Texas at Austin, McDonald Observatory, McDonald Observatory, TX 79734, USA}

\begin{abstract}
We map the distribution and properties of the Milky Way's interstellar medium as traced by diffuse interstellar bands (DIBs) detected in near-infrared stellar spectra from the SDSS-III/APOGEE survey.
Focusing exclusively on the strongest DIB in the $H$-band, at $\lambda \sim 1.527 \,\mu{\rm m}$, we present a projected map of the DIB absorption field in the Galactic plane, using a set of about 60,000 sightlines that reach up to 15 kpc from the Sun and probe up to 30 magnitudes of visual extinction. The strength of this DIB is linearly correlated with dust reddening over three orders of magnitude in both DIB equivalent width ($W_{\rm DIB}$)
and extinction, with a power law index of $1.01 \pm 0.01$,
a mean relationship of $W_{\rm DIB}$/$A_V = 0.1$~\AA~mag$^{-1}$ and a dispersion of $\sim$0.05~\AA~mag$^{-1}$ at extinctions characteristic
of the Galactic midplane. 
These properties establish this DIB as a powerful, independent probe of dust extinction over a wide range of $A_V$ values.
The subset of about 14,000 robustly detected DIB features have an exponential $W_{\rm DIB}$ distribution.
We empirically determine the intrinsic rest wavelength of this transition to be $\lambda_0 = 15\,272.42$~\AA\, and then calculate absolute radial velocities of the carrier, which display the kinematical signature of the rotating Galactic disk. We probe the DIB carrier distribution in three dimensions and show that it can be characterized by an exponential disk model with a scaleheight of about 100 pc and a scalelength of about 5 kpc. Finally, we show that the DIB distribution also traces large-scale Galactic structures, including the central long bar and the warp of the outer disk.
\end{abstract}


\section{Introduction} \label{sec:intro}

The diffuse interstellar bands (DIBs) are a set of a few hundred 
optical-infrared (IR) absorption features observed with line widths broader than typical atomic interstellar absorption lines \citep[e.g.,][]{Herbig_1975_dibs,Herbig_1995_dibs,Hobbs_2008_DIBcatalog}. The features were first reported in the 1920s \citep[e.g.,][]{Heger_1922_stationarylines} and identified as interstellar in origin in the 1930s \citep[e.g.,][]{Merrill_1934_dibs,Merrill_1938_dibs}. Despite nearly a century of research, the carriers responsible for these absorption lines are still unknown. Plausible candidates have included dust grains \citep[but see][]{Cox_2007_dibpolarization,Cox_2011_dibpolarization,Xiang_2011_dibs-vs-2175bump}, polycyclic aromatic hydrocarbons \citep[PAHs; e.g.,][]{Crawford_1985_pahdibs,Leger_1985_pahdibs,vanderzwet_1985_pahdibs,Salama_1999_pahdibs}, 
small linear molecules \citep[e.g.,][]{Oka_2013_Herschel36DIBs},
and fullerenes and fullerene-related molecules \citep[such as C$_{60}$; e.g.,][]{Foing_1994_C60dibs,Iglesias-Groth_2007_fullerenes4430DIB,GarciaHernandez_2013_fullerenedibs,Cami_2014_DIBfullerenes}. 
However, none of these species yet have both the observational and laboratory evidence to be associated definitively with any particular DIB.

Most of the DIBs currently known are at visible wavelengths. Observationally, their study has relied primarily on analyses of limited sets of nearby early-type stars whose nearly featureless spectra allow for a straightforward detection of absorption lines from the intervening interstellar medium (ISM). Recently, a few teams have made use of large sky surveys to detect and map out DIBs.
Using the 
Sloan Digital Sky Survey (SDSS), \citet{Yuan_2012_sdssdibs} reported the detection of optical DIBs in about 2000 medium-resolution ($R\sim2000)$ stellar spectra, 
and \citet{Lan_2014_sdssDIBs} detected high-latitude DIBs from composite SDSS spectra of stars, quasars, and galaxies. \citet{Kos_2013_RAVEdibs} measured optical DIBs in several thousand composite stellar spectra, each stacked from dozens of individual sightlines from the 
RAdial Velocity Experiment \citep[RAVE;][]{Steinmetz_2006_raveDR1}.

\begin{figure}[t]
   \includegraphics[trim=3.7in 1.0in 1.0in 5.5in, clip, angle=90, width=0.5\textwidth]{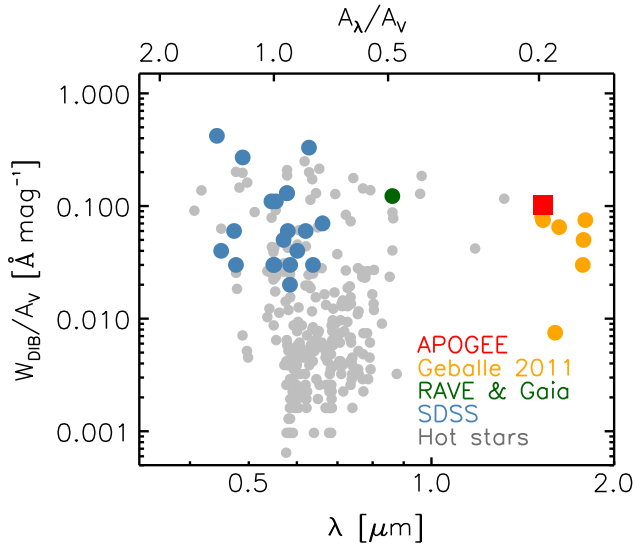} 
\caption{
Distribution of wavelength ($\lambda$) and mean equivalent width-to-extinction ($W_{\rm DIB}$/$A_V$) ratios for 273 optical and NIR DIBs.
Along the top axis, wavelength has been converted to approximate 
extinction relative to $V$-band.
The red square indicates the DIB studied here, using the $W_{\rm DIB}$/$A_V$ ratio derived in Section~\ref{sec:red} and the rest wavelength derived in Section~\ref{sec:kinematics}. The gray points are from the compilation in \citet{Jenniskens_1994_dibs}, and the orange points are the other $H$-band DIBs reported by \citet{Geballe_2011_IRdibs}. Green indicates the DIB within the RAVE and Gaia wavelength ranges \citep{Kos_2013_RAVEdibs}, and blue points are measurements of DIBs in SDSS stellar spectra \citep{Lan_2014_sdssDIBs}.
}
\label{fig:DIBs_w-to-av}
\end{figure}

The first IR DIBs were identified in the $J$-band \citep{Joblin_1990_IRDIBS}. The first detections of longer wavelength $H$-band features were reported by \cite{Geballe_2011_IRdibs}. Since then, a handful of other IR features have been identified \citep[e.g.,][]{Cox_2014_xshooterDIBs}. Figure~\ref{fig:DIBs_w-to-av} shows the wavelengths and relative strengths of a compilation of known DIBs \citep[from the updated catalog\footnote{\url{http://leonid.arc.nasa.gov/DIBcatalog.html}; accessed 22 May 2013. 
The \citet{Geballe_2011_IRdibs} $H$-band DIBs shown exclude those whose $W_\lambda$ measurements include multiple adjacent features. The $W$-to-extinction ratio is computed using $A_V=20$~mag as estimated in their paper.} of][see also \cite{Hobbs_2008_DIBcatalog,Hobbs_2009_DIBcatalog}]{Jenniskens_1994_dibs,Geballe_2011_IRdibs}. 
The red square indicates the DIB analyzed in this paper.

While typical DIB studies have focused on obtaining constraints on the nature of the DIB carriers (for example, through correlations with each other or with their line-of-sight environmental properties), we show in this paper that it is possible to use these absorption features effectively to probe Galactic structure, despite not knowing the identity of the carriers. 
Steps towards this approach, including the spatial mapping of DIB strengths and the assessment of spatial correlations, have been explored by, e.g., 
\citet{vanLoon_2009_omegaCenDIBs,vanLoon_2013_30dorDIBs} in their studies of the ISM foreground to $\omega$~Cen and of the LMC's Tarantula Nebula,
and by \citet{Kos_2014_RAVEdibmaps} in their integrated absorption map of optical DIBs out of the Galactic midplane.

Expanding this approach to map the DIBs efficiently and directly on large scales within the bulk of the Milky Way's ISM
requires (1) the use of a relatively strong DIB to maximize sensitivity, as well as (2) the ability to probe a broad range of Galactic environments --- including making use of stars located behind up to 30 magnitudes of visual extinction (i.e., towards the Galactic bulge). 
These goals can be achieved using data from the Apache Point Observatory Galactic Evolution Experiment \citep[APOGEE;][]{Majewski_2012_apogee}, 
a large near-IR spectroscopic survey spanning about 100,000 stars
in a wide range of Galactic environments.
NIR observations allow us to probe heavily reddened stars --- such as those in the Milky Way bulge and midplane, wherein lies the majority of the ISM and stellar mass ---  at sufficient spectral resolution and S/N to identify interstellar absorption features.  

We present a census of the $\lambda \sim 1.527~\mu$m DIB, which we detect in tens of thousands of individual sightlines observed by APOGEE. As shown in Figure~\ref{fig:DIBs_w-to-av}, this feature is extremely well-suited to probing the distant ISM because it is both strong (relative to the dust content) and at a wavelength where dust extinction (in magnitudes) is about six times weaker than at optical wavelengths. Our large number of sightlines is the final crucial piece for this approach of using DIBs 
to trace diffuse Galactic metals and Galactic structure,
as well as
to constrain the global environmental properties of the DIB carriers themselves.

The outline of the paper is as follows: we present the APOGEE data and flux residual selection in Section~\ref{sec:data}. We discuss results from the mean DIB absorption field in Section~\ref{sec:mean_properties}, including the relationship with reddening, and results from the properties of well-detected features in Section~\ref{sec:feature_properties}, including a determination of the rest-frame wavelength of the feature. Finally, in Section~\ref{sec:3dmap} we illustrate the use of the selected DIB to reveal ISM features in the structure of the Milky Way.

\section{Data} 
\label{sec:data}


\subsection{The APOGEE Survey} \label{sec:apogee}

\begin{figure}[!hptb]
\includegraphics[angle=90, trim=1.3in 4.5in 1.5in 1.4in, clip, width=0.5\textwidth]{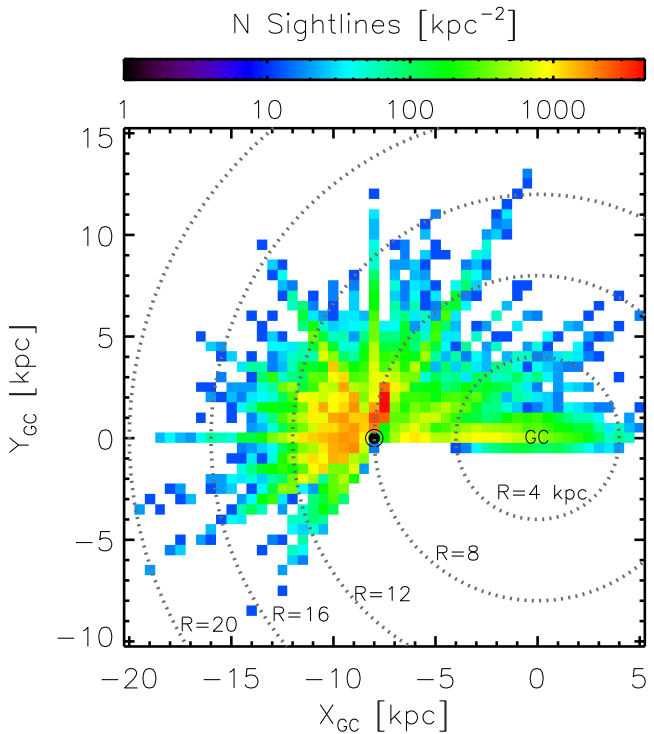}
\caption{
Surface density (kpc$^{-2}$) in the Galactic disk of APOGEE stars in our subsample used in this paper..
The label ``GC'' indicates the Galactic Center, 
and the circled dot is at the position of the Sun (assuming $R_{{\rm GC, \sun}} = 8$~kpc).
The dotted circles have constant Galactocentric distances of $R_{\rm GC} = 4$, 8, 12, 16, and 20~kpc.
}
\label{fig:xymap_nstars}
\end{figure}

APOGEE, part of the Sloan Digital Sky Survey III \citep[SDSS-III;][]{Eisenstein_11_sdss3overview}, is a high-resolution ($R \sim 22,500$), $H$-band ($\lambda = 1.51-1.70~\mu$m) spectroscopic survey targeting just over 10$^5$ red giant candidate stars spread throughout the Milky Way's halo, disk, and bulge \citep{Majewski_2012_apogee,Zasowski_2013_apogeetargeting}.  
Figure~\ref{fig:xymap_nstars} shows the surface density of stars used in this paper,
in a face-on projection of the Galactic disk (seen from the North Galactic Pole) with the Sun at $(X_{\rm GC}, Y_{\rm GC}, Z_{\rm GC}) = (-8, 0, 0)$~kpc.  The dotted lines of constant Galactocentric radius
highlight the span of APOGEE stars over a large fraction of the disk.

The survey uses a custom-built spectrograph \citep{Wilson_2012_apogee}, attached via a fiber optic train to the 2.5-m Sloan telescope \citep{Gunn_2006_sloantelescope} at the Apache Point Observatory, and is capable of observing 300 targets simultaneously with standard SDSS plugplates. The spectra are reduced and corrected for telluric absorption and airglow emission using APOGEE's custom data reduction pipeline (D.~L.~Nidever et al., in preparation).  This pipeline also measures radial velocities (RVs) for the stars, with a typical uncertainty of $\lesssim$150~m~s$^{-1}$. In this work we make use of APOGEE data from the first two years of the survey, using version v402 of the combined reduction+analysis 
pipeline\footnote{Part of these data are currently available as part
of the SDSS Data Release 10 (DR10; \url{www.sdss3.org/dr10/data\_access/}), and the
remainder will be available in DR12, to be released by the end of 2014.}.

\subsection{Interstellar Spectrum Estimation}
\label{sec:aspcap}

APOGEE's target sample comprises mostly cool K and M giants ($T_{\rm eff} \sim 3500-5000$~K), whose spectra ($F_\lambda$) are dominated by stellar atmospheric absorption lines.  Reliable removal of these features is crucial for identifying interstellar absorption lines. To do this,
we make use of the APOGEE Stellar Parameters and Chemical Abundances Pipeline \citep[ASPCAP;][and in preparation]{GarciaPerez_2014_aspcapaas}. For each star, ASPCAP provides a best-fit model spectrum $F'_\lambda$, 
estimated from a large grid of synthetic spectra \citep[derived for pre-computed model atmospheres;][]{Meszaros_2012_APOGEEatmogrids} using multi-dimensional $\chi^2$-minimization. 
We use these best-fitting synthetic spectra as templates for the stellar contribution to the total observed spectral energy distribution (SED):
\begin{equation}
R_\lambda = \frac{F_\lambda}{ F'_{\lambda}},
\end{equation}
which allows us to search for interstellar absorption lines in the residual spectrum, $R_\lambda$. 
We remove any large-scale fluctuations of the residuals around unity by subtracting a median estimate of $R_\lambda$ using a width of $\sim$20~\AA\,
(100 pixels), with a 40~\AA-wide mask applied at the expected location of the DIB.

\begin{figure*}[!htbp]
\begin{center}
  \includegraphics[trim=0.8in 1.2in 1in 1.33in, clip, width=\textwidth]{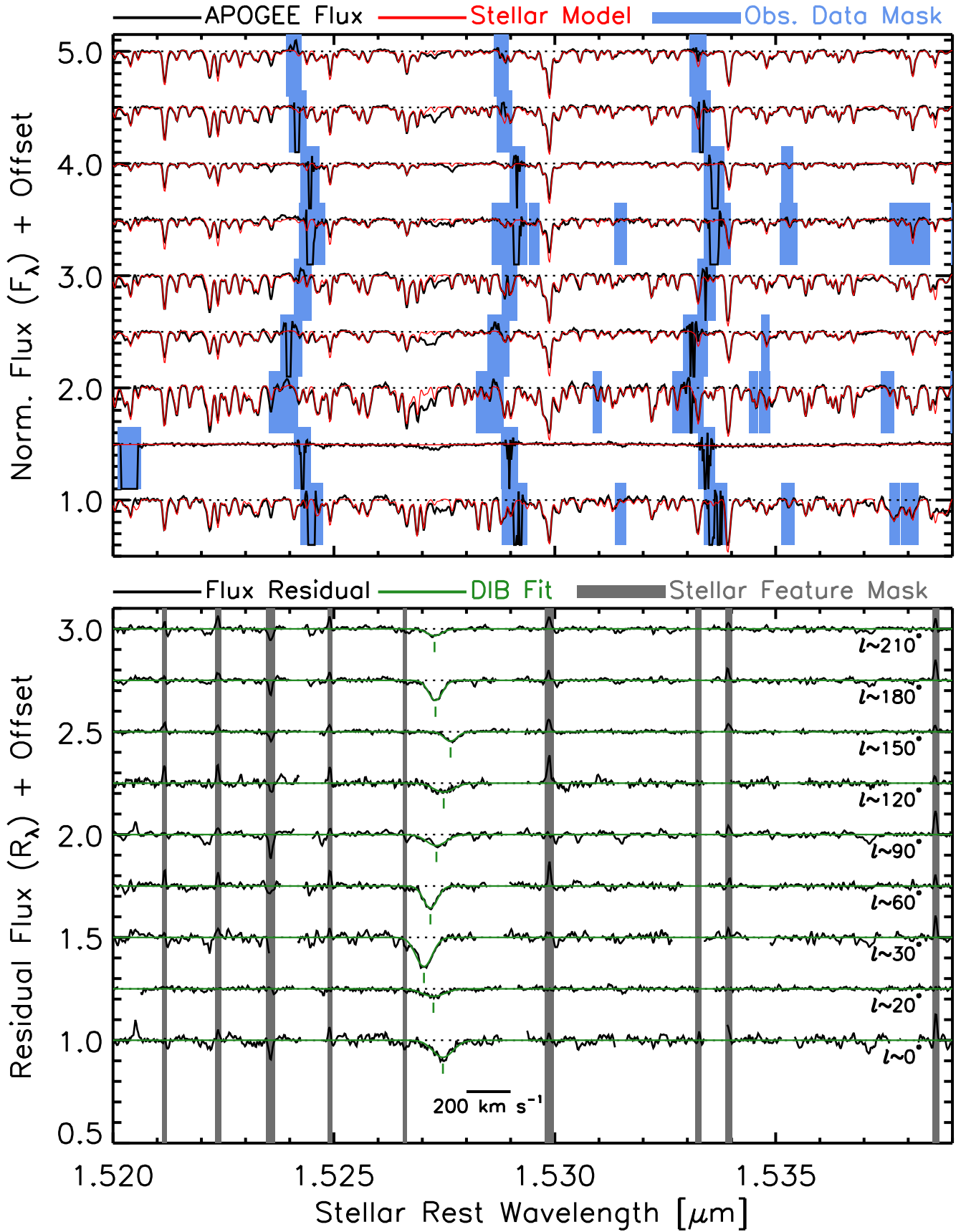} 
\caption{
Examples of APOGEE spectra and the removal of the stellar contributions to the spectra.  
The top panel contains the continuum-normalized flux of nine observed stellar sightlines (${\rm S/N} \gtrsim 110$; black lines), 
spanning 190~\AA\; of the APOGEE wavelength range 
(range specified at the bottom of the figure).  
Overplotted in red are the best-fit synthetic spectra,
and the blue shading indicates regions significantly affected by sky
emission or other sources of uncertainty.
The bottom panel contains these sightlines' model-normalized residuals $R_\lambda$ 
(black lines), 
overplotted with the best-fit DIB profiles in green (Section~\ref{sec:residuals}). The shaded spans above have been masked.  The gray shaded columns indicate regions affected by persistent stellar residuals.
The spectra are ordered by Galactic longitude, and the impact of the rotation
of the Galactic ISM can be seen in the shifting of the observed DIB central wavelengths
(indicated with vertical green lines; Section~\ref{sec:kinematics}).
The horizontal black bar below the bottom residual shows the velocity scale.
}
\label{fig:aspcap_fit}
\end{center}
\end{figure*}

As we focus only on the DIB feature around $\lambda=1.527\mu$m, in this paper we consider the quantity $R_\lambda$ over the 
range $1.515\,\mu \rm m \le \lambda \le 1.560\,\mu \rm m$. Some examples of the construction of the residuals is shown in Figure~\ref{fig:aspcap_fit}, for a $\Delta \lambda = 190$~\AA\, span of APOGEE's shortest-wavelength detector. The top panel contains the continuum-flattened flux $F_\lambda$ for eight typical giant stars and one ``hot star'' telluric calibrator, shown in black. 
The giant stars span a range of temperatures and metallicities: 
$4020 \lesssim T_{\rm eff} \lesssim 5090$~K and $-0.5 \lesssim {\rm [Fe/H]} \lesssim +0.2$ (all parameters given in Table~\ref{tab:aspcap_fit}).
The observed flux has been overplotted in red with the spectrum's best-fitting synthetic model $F^\prime_\lambda$, as determined by ASPCAP.
Also shown in blue are regions of each spectrum with high flux uncertainties
(for example, caused by insufficiently subtracted sky emission).
In the bottom panel, which is discussed in greater detail in Section~\ref{sec:residuals},
we show these stars' flux residuals $R_\lambda$ as black lines,
where the potentially problematic pixels have been masked.
These are overplotted with the Gaussian profiles that best fit the residual features
(green).  
Shown here in gray are the wavelengths of frequently ill-fit stellar lines, which are also masked in the profile fits.
All spectra are presented in the stellar rest frame.

\setlength{\tabcolsep}{0.07cm}
\begin{deluxetable}{lcccc|ccc}[!hptb]
\tablewidth{0pt}
\tablecolumns{8}
\tablecaption{Stellar and DIB Parameters in Sample ASPCAP Fits (Figure~\ref{fig:aspcap_fit})}
\tablehead{ \colhead{} & \multicolumn{4}{c}{Stellar Parameters} & \multicolumn{3}{c}{DIB Parameters} \\ 
  \vspace{-5pt} \\
                \colhead{} & \colhead{$T_{\rm eff}$} & \colhead{$\log{g}$}  & \colhead{[Fe/H]} & \colhead{$A_{\rm V}$} & \colhead{$W_{\rm DIB}$} & \colhead{$\lambda_{\rm DIB}$} & \colhead{$\sigma_{\rm DIB}$} \\
                 \colhead{2MASS ID} & \colhead{[K]} & \colhead{} & \colhead{} & \colhead{[mag]} & \colhead{[\AA]} & \colhead{[\AA]} & \colhead{[\AA]}}
\startdata
06441471+0400302 & 4934 & 3.0 & $+0.10$ & 1.54 & 0.14 & 15\,272.8 & 1.71 \\
05414091+2857412 & 4426 & 2.0 & $-0.25$ & 2.49 & 0.38 & 15\,273.0 & 1.55 \\
03583582+5246569 & 4938 & 3.1 & $-0.48$ & 1.43 & 0.17 & 15\,276.4 & 1.49 \\
00140182+6233305 & 5085 & 3.1 & $-0.03$ & 1.57 & 0.25 & 15\,274.9 & 2.30 \\
21175329+4736271 & 4279 & 1.8 & $-0.09$ & 2.74 & 0.30 & 15\,273.2 & 2.23 \\
19420307+2317393 & 4984 & 3.0 & $-0.05$ & 3.91 & 0.42 & 15\,271.8 & 1.58 \\
18491800$-$0153008 & 4460 & 2.2 & $+0.23$ & 8.64 & 0.66 & 15\,270.4 & 1.86 \\
18244326$-$1210496\tablenotemark{a} & --- & --- & --- & 0 & 0.24 & 15\,272.5 & 2.54 \\
17531202$-$2921172 & 4017 & 1.2 & $-0.42$ & 3.74 & 0.48 & 15\,274.7 & 2.31 
\enddata
\tablecomments{Stars are ordered top to bottom as in Figure~\ref{fig:aspcap_fit}.  The stellar parameters 
are from version v402 of APOGEE's reduction+ASPCAP pipeline
and may differ from those in the public releases.}
\tablenotetext{1}{This is an early type star with parameters beyond ASPCAP's spectral library.  
The RJCE-derived $A_{\rm V}$ is consistent with zero within the photometric uncertainties.}
\label{tab:aspcap_fit}
\end{deluxetable}

The APOGEE dataset we are currently using provides us with observations of 96,938 unique stars, all of which have an ASPCAP best fit estimate of $F'_\lambda$. The quality of these estimates varies: we typically find the standard deviation in $R_\lambda$ to be on the order of a few percent, but in some cases this value can be substantially greater. This can be due to a number of effects, including a poor ASPCAP fit, insufficiently subtracted sky emission lines, poor calibration of telluric absorption lines, etc. For our analysis we define a sample of ``clean'' residuals whose fluctuations are sufficiently small and well-understood. We define this sample with the following requirements:
\begin{itemize}
\item locally good ASPCAP fit: we keep only those systems for which the variance in $R_\lambda$ is lower than the variance in the flux, which indicates that the ASPCAP synthetic spectrum captures at least a fraction of the stellar flux variations. 
Visual inspection of many spectra indicate that a ratio value of about 0.55 is appropriate. We therefore use $\sigma(R_\lambda)/\sigma(F_\lambda) \leq 0.55$ as a limit.
We note that the exact threshold value is not critical in our analysis because we are not aiming at maximizing the completeness of our line detection;
\item ``smooth'' residual: we require the standard deviation of $R_\lambda$ to be less than 5\%, which indicates that deviations due to an insufficient correction of atmospheric emission and/or absorption lines are globally negligible; and
\item well measured stellar RV: poor velocity estimates and/or the presence of stellar companions can significantly affect the final co-added spectrum obtained from the combination of different epochs. To remove these effects, we require that if a spectrum is a composite of multiple epoch observations (i.e., ${\rm NVISITS} \neq 1$), it have no significant RV variation between epochs (defined as ${\rm VSCATTER} \le 1$~km~s$^{-1}$).
\end{itemize}
After applying these selection criteria, we obtain a sample of 58,605 ``clean'' $R_\lambda$ residuals, 
corresponding to about 60\% of the parent sample. 
We note that we do not include any restrictions on the ASPCAP global fitting metrics or derived stellar parameters themselves. 
Because we are focusing on just a fraction of the spectral range, 
these requirements serve to identify sightlines with clean $R_\lambda$ residuals around the DIB feature, 
regardless of how well the global spectrum is reproduced.

\subsection{DIB Characterization} 
\label{sec:residuals}

To characterize any DIB absorption present along each line of sight, we perform a Gaussian fit in the full range of $R_\lambda$
($1.515\,\mu \rm m \le \lambda \le 1.560\,\mu \rm m$), leaving the profile central wavelength as a free parameter.
The validity of this approach, however, requires that the flux residuals $R_\lambda$ be centered around unity and 
that no other absorption or emission lines are present. 
To do so, for each star we must mask a number of features:

(1) Telluric absorption and airglow features that are imperfectly corrected by the data reduction pipeline.  Of these, the airglow residuals tend to dominate.
These features can be easily identified because they always appear at the same wavelength in the observer frame, and they rarely affect ASPCAP's global model fitting because the affected wavelength ranges are both sparse and downweighted in the ASPCAP fitting algorithm. To mask the telluric and airglow features from the DIB fit, we use the error spectrum provided by the data reduction pipeline, supplemented by an additional mask flagging pixels where the sky flux correction is $\ge$25\% of the total observed flux. 
These pixels are indicated by blue shading in the top panel of Figure~\ref{fig:aspcap_fit}.

(2) Stellar lines that are improperly fit by ASPCAP.  This category can include overall poor ASPCAP fits as well as cases where the global fit is good but a small number of observed lines are not present in the line list used in generating the synthetic spectra grid.  The relatively under-explored nature of high resolution $H$-band stellar spectroscopy necessitated much effort to create a custom line list for APOGEE (M.~Shetrone et al., in preparation). Previously-unknown transitions have been identified in APOGEE spectra and added to the line list (e.g., Nd; S.~Hasselquist et al., in preparation), but unidentified lines remain.  These features can be identified by their consistent wavelength in the stellar rest frame, and they are far too sparse and weak to impact ASPCAP's ability to fit the rest of the spectrum.  We identify systematic ASPCAP residuals using a median residual spectrum of all stars with high S/N ($\ge$100 per pixel) and $|b| \ge 60^\circ$.  The mean of this median $R_\lambda$ residual is 0.999, with a standard deviation of 0.004, and pixels deviating from the mean by more than four standard deviations are masked out in all of the individual $R_\lambda$ spectra.
These pixels are indicated by gray shading in the bottom panel of Figure~\ref{fig:aspcap_fit}.
These criteria provide us with a spectral mask for each star that is used when performing a Gaussian fit, 
resulting in a best-fit peak amplitude $A$, central wavelength $\lambda$, and width (standard deviation) $\sigma$. 
The equivalent width $W$ of the best fit line is given by
\begin{eqnarray}
W &=& \int_{\lambda_1}^{\lambda_2} (1-R_\lambda)\; {\rm d}\lambda \nonumber\\
&=&\sqrt{2\pi}\,A\,\sigma .
\end{eqnarray}
We note that, while some optical DIBs show clear asymmetries that preclude a single Gaussian fit \citep[due to intrinsic asymmetries, blending with stellar or other interstellar lines, or local ISM properties; e.g.,][]{Jenniskens_1994_dibs,Dahlstrom_2013_herschel36dibs,Kos_2013_dibs-physconditions}, this approach is found to be appropriate for the present analysis.
The bottom panel of Figure~\ref{fig:aspcap_fit} shows some example DIB feature fits,
with the corresponding fit parameters ($W$, $\lambda$, and $\sigma$) given in Table~\ref{tab:aspcap_fit}.
The spectra are shown in the stellar rest frame, 
but the excess absorption due to the DIB appears at different wavelengths, 
highlighted by the vertical green lines.
The spectra are ordered by Galactic longitude, and the shifting of the observed DIB central wavelengths is due to the rotation of the Galactic ISM (see Section~\ref{sec:kinematics}).

\begin{figure*}[h]
\includegraphics[angle=90, trim=3.5in 0.8in 1in 0.4in, clip, width=\textwidth]{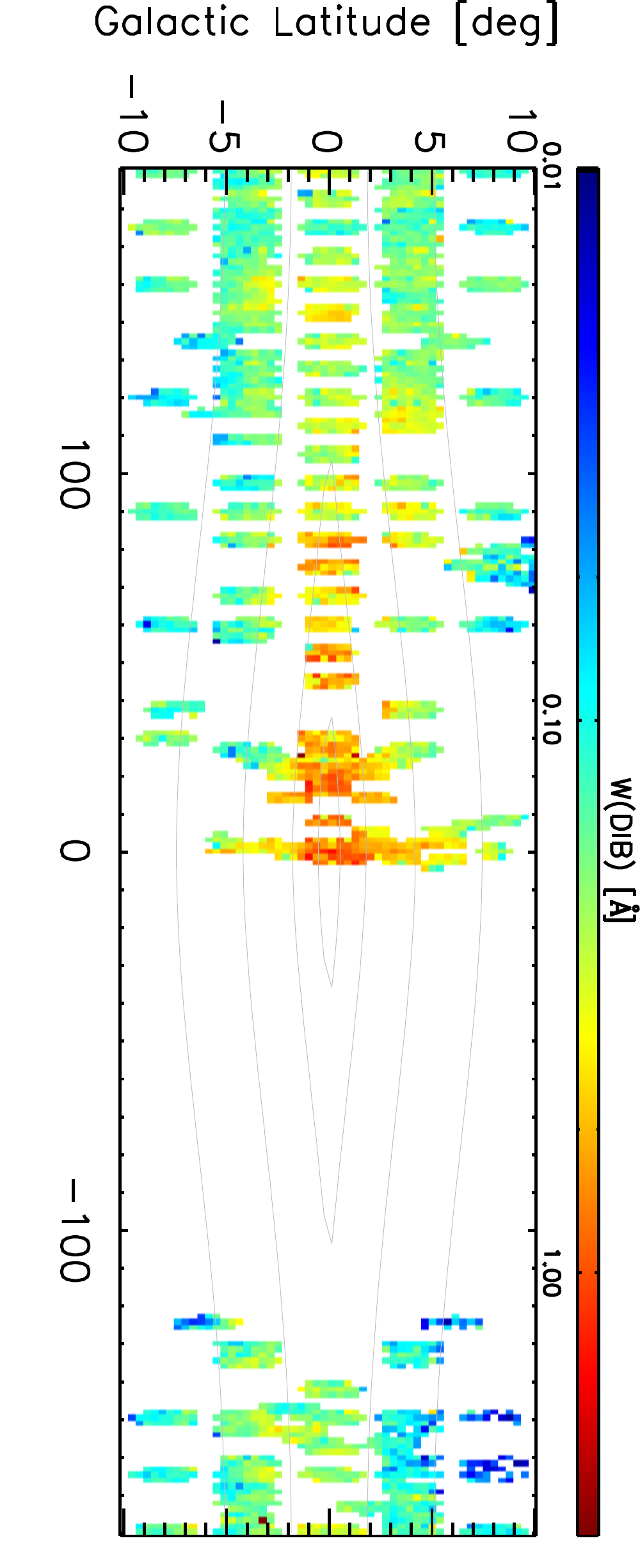} 
\includegraphics[angle=90, trim=3.5in 0.8in 1in 0.4in, clip, width=\textwidth]{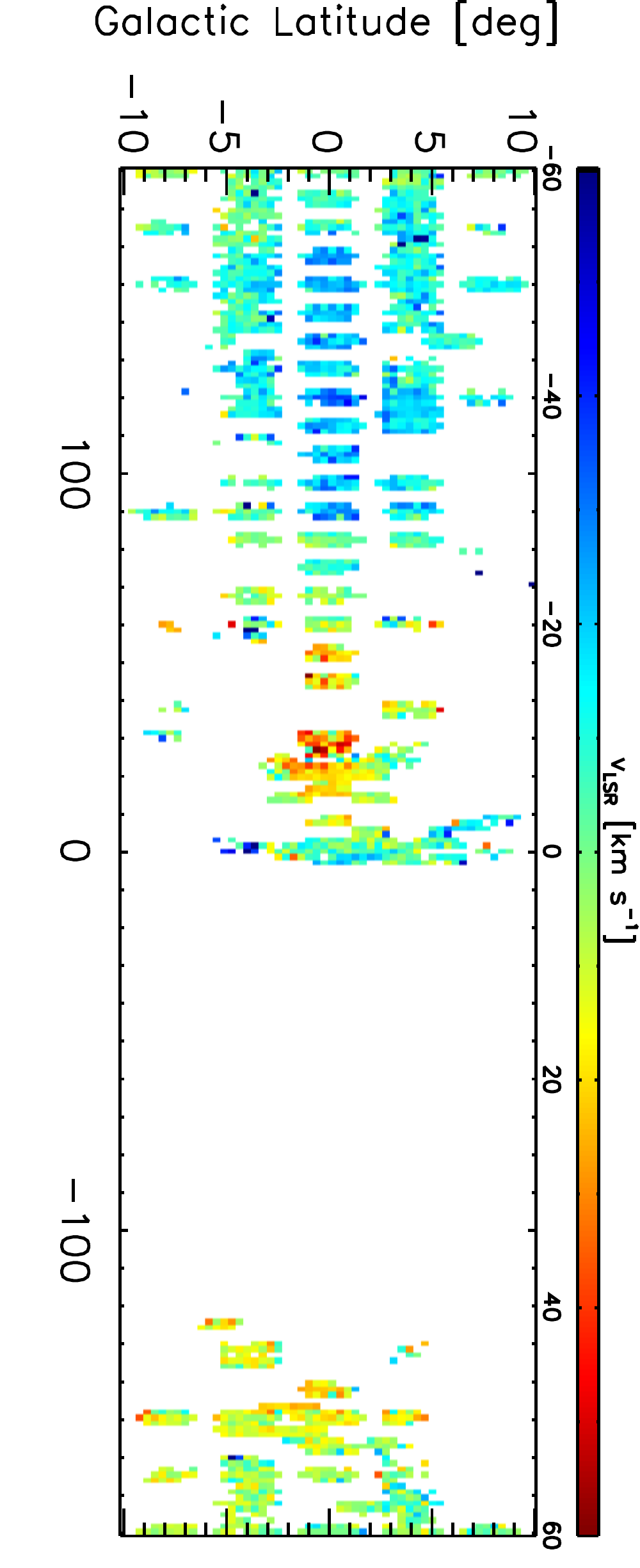}
\includegraphics[angle=90, trim=3.5in 0.8in 1in 0.4in, clip, width=\textwidth]{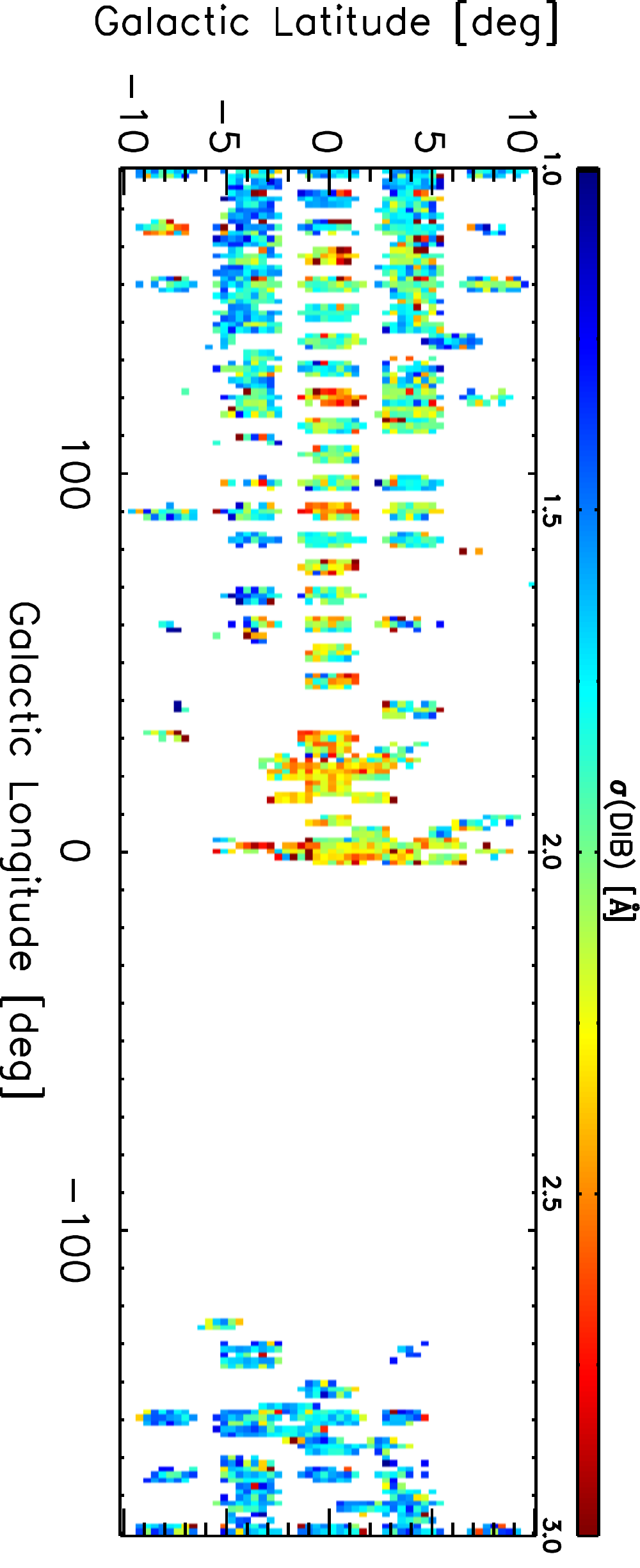}
\caption{
{\it Top}: Median $W(l,b)$ of the DIB absorption field, measured from all clean residuals, in 0.8~deg$^2$ pixels (Section~\ref{sec:absorption_field}).  The gray contours outline a projected disk distribution at 6~kpc (Section~\ref{sec:3dmap}).
{\it Middle}: Median velocity ($v_{\rm LSR}$) of well-detected DIB features (Section~\ref{sec:kinematics}).
{\it Bottom}: Median profile width $\sigma$ of well-detected DIB features (Section~\ref{sec:fwhm}).  A profile width of 2~\AA\, corresponds to about 40~km~s$^{-1}$ here.
All panels share a common x-axis, shown below the bottom panel.
}
\label{fig:param_map}
\end{figure*}

\section{Mean Properties of the DIB Absorption Field}
\label{sec:mean_properties}

\subsection{The DIB Absorption Field} 
\label{sec:absorption_field}

Due to the noise properties of the flux residuals $R_\lambda$, and the expected broad range of possible DIB strength, we note that the best fit amplitude $A$ (or, similarly, the equivalent width $W$) is in general a noise-dominated quantity and can thus be either positive or negative on an object-by-object basis. In the low DIB S/N regime, though we cannot meaningfully consider the properties of individual absorber systems, we can nevertheless extract useful signals by considering average values across the DIB absorption field. This is illustrated in Figure~\ref{fig:param_map}, where the top panel shows the distribution of median values of the DIB equivalent width, $W_{\rm DIB}$, as a function of Galactic coordinates. Each pixel spans 0.8~deg$^2$,
and the measured values are typically averaged over 5--25 stars. This approach allows us to detect the DIB feature over three orders of magnitude in equivalent width, from about 0.01 to a few \AA. It demonstrates that diffuse interstellar bands can be used as an ISM tracer to map out coherent structure of the Milky Way --- not only large-scale gradients with Galactic
longitude and latitude, but also distinct clouds and patterns, such as the one at 
$(l,b) = (150^\circ,-3.5^\circ)$, which appears to extend up into the midplane at $l=140^\circ$.

\subsection{Linear Correlation With Dust Extinction} 
\label{sec:red}

\begin{figure*}[ht]
\begin{center}
   \includegraphics[trim=4.3in 5.5in 0.8in 1.1in, clip, angle=90, width=0.8\textwidth]{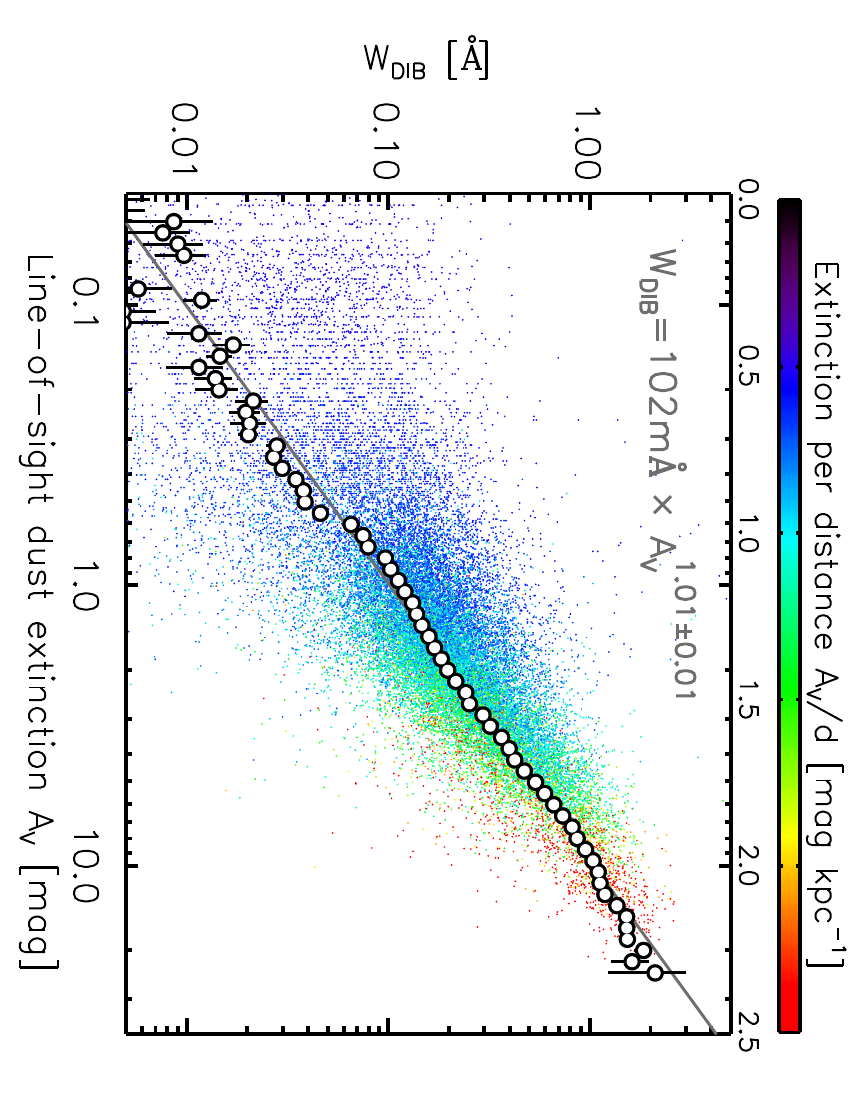} 
\caption{The equivalent width $W_{\rm DIB}$ as a function of foreground extinction $A_V$.  
Small points are the individual sightline measurements, with the color indicating
the mean extinction per kpc along each sightline towards the background star.
The larger white circles are the $W_{\rm DIB}$ medians in each $\Delta\log{A_V}=0.04$ bin, along with the uncertainty in the median,
and the gray line is the fitted $W_{\rm DIB}/A_V$ relationship.
Note that at low $A_V$, a large fraction of the uncertainty is coming from the extinction measurement, rather than the DIB equivalent width.
}
\label{fig:ew_vs_extinction}
\end{center}
\end{figure*}

Many DIBs are known to present a high degree of correlation with dust extinction. We can make use of our measured (noise dominated) DIB absorption strength along all lines of sight defined above to assess this correlation in both strong and weak DIB systems.
By averaging over a sufficiently large number of systems, one can probe low-level DIB features weaker than the detection threshold given by the noise level of individual spectra.

We calculate the line-of-sight dust extinction $A_V$ for each star with the Rayleigh-Jeans Color Excess method \citep[RJCE;][]{Majewski_11_RJCE}, which uses the fact that normal stars generally have identical intrinsic colors in NIR+MIR filter combinations that span the peak and Rayleigh-Jeans tail of the stellar SED. This homogeneity means that redder colors observed at these wavelengths can be unambiguously attributed to reddening by dust. 
We emphasize that this approach also ensures that each extinction value used arises from the same integrated sightline that gives rise to the observed DIB feature, as opposed to estimates from integrated extinction maps or those requiring assumptions about the background source distance, which may probe vastly different sightline lengths.

Figure~\ref{fig:ew_vs_extinction} shows the measured $W_{\rm DIB}$ as a function of line-of-sight $V$-band extinction towards the background 
stars\footnote{Note that negative $W$ values are not shown, given the choice of a logarithmic $y$-axis, but they are included in the medians.} (small points), 
along with the median value of the DIB equivalent width (white circles).  
The correlation, measured over three orders of magnitude in both $A_V$ and $W_{\rm DIB}$,
is extremely linear; a power-law fit of the form
\begin{equation}
W_{\rm DIB} \propto A_V^\alpha
\end{equation}
results in:
\begin{equation} \label{equ:ew_vs_extinction}
\begin{aligned}
{\rm \normalfont amplitude}~W_{\rm DIB}/A_V &= 102 \pm 1~{\rm m\normalfont{\AA}~mag}^{-1} \\
{\rm \normalfont index}~\alpha &= 1.01 \pm 0.01.
\end{aligned}
\end{equation}
The fact that the power-law index $\alpha$ is consistent with unity over such a large dynamic range of extinction
is remarkable, as well as significant in that it demonstrates this DIB to be a highly effective tracer of interstellar dust,
even in very dense, dusty regions (see below).

Furthermore,
the level of DIB absorption per unit dust extinction for this feature is rather high compared to other DIBs (e.g., Figure~\ref{fig:DIBs_w-to-av}). 
The mean $W_{\rm DIB}/A_V$ relationship observed for all features in our sample is $102.2 \pm 0.5~{\rm m\normalfont\AA~mag}^{-1}$,
consistent with the amplitude of the power-law fit in Equation~\ref{equ:ew_vs_extinction}.
The total dispersion $\sigma_{W/A}$ is about 107~m\AA~mag$^{-1}$, 
which is dominated by the noise at $A_V \lesssim 1$;
the dispersion at $A_V \ge 1.6$~mag, which characterizes nearly all of the Galaxy midplane inside the solar circle 
\citep[e.g.,][]{Nidever_2012_rjcemaps}, is only 50~m\AA~mag$^{-1}$ (i.e., $\sigma_{W/A}/(W/A) = 0.5$). 

The strength of this feature and the tight linearity of its correlation with dust extinction thus make it a powerful, independent indicator of foreground extinction, even when photometric or other methods are insufficient --- for example, (1) in low-$|b|$ regions where dust emission-based maps are unreliable, (2) towards particularly cool or unusual stars where color excess methods fail, or (3) for stars with unknown distances where 2- and 3-D extinction maps may be ambiguous \citep[e.g.,][]{Schultheis_14_2-3Dextinction}. These foreground extinction values can be very important for stellar distance calculations that employ stellar models, when the intrinsic colors are not known a priori.

Notably, the observed correlation ranges from strong features in the spectra of stars with $A_V \sim 26$~mag to weak features in stars with $A_V \sim 0.04$~mag, where individual measurements of both $W$ and $A_V$ are in noisy regimes.  As the median trend demonstrates, 
the relationship observed at high $W$ and $A_V$ extends consistently to weaker features, past the noise floor induced by our purer selection of well-detected features
(Section~\ref{sec:catalog}).
The apparent deviations from the mean trend around $A_V \lesssim 0.8$~mag 
are driven by increased noise (e.g., larger numbers of noise-dominated $W_{\rm DIB}$ values measured in low-$A_V$ sightlines, 
fractionally larger extinction uncertainties) and smaller counts in the bins.  
These sightlines may also be more strongly impacted by the effects of single dark clouds or local radiation fields 
\citep[e.g.,][]{Raimond_2012_extinctionDIBs,Chen_2013_coolstarDIBs}, 
whereas the high-reddening sightlines are more likely to represent integration through multiple clouds or fields.
We note that the few $W_{\rm DIB}$ medians deviating from the trend at high $A_V$ represent small counts with large dispersions, 
and their departure from the best-fit trend is only marginally significant.

The color of the 
points\footnote{The individual points shown are a subset of the total sample of clean residuals for which we have
reliable distances, as detailed in Section~\ref{sec:3dmap}, but the mean relationship
and dispersions discussed above are derived from {\it all} clean residuals, 
as described in Section~\ref{sec:absorption_field}.}
in Figure~\ref{fig:ew_vs_extinction} denotes each sightline's mean extinction per unit distance, $A_V/d$ (mag~kpc$^{-1}$).  
This quantity is a rough proxy for mean dust density along the line of sight; 
$A_V \sim 0.9$~mag~kpc$^{-1}$ is an approximate limit between  ``diffuse'' and ``dense'' ISM environments \citep[e.g.,][]{Clayton_2014_dibsndust}. 
We see no significant
decrease in the $W_{\rm DIB}$/$A_V$ ratio at higher line-of-sight mean densities,
which is qualitatively different from the tendency for many optical DIBs to stop increasing in strength,
or even grow weaker, in sightlines probing dense, high-extinction clouds.

This weakening behavior may be considered evidence 
for DIB carrier depletion out of the gaseous ISM
onto dust grain surfaces in dense clouds, 
the coating of the carrier particles with ice mantles in the clouds,
or else the clouds providing sufficient shielding
to block the radiation necessary to produce an ionized DIB carrier
\citep[e.g.,][]{Snow_1974_densecloudDIBs,Krelowski_1992_molecules-vs-DIBs,Cami_1997_singlecloudDIBs,Kos_2013_dibs-physconditions}.
The $W_{\rm DIB}$/$A_V$ ratio for this DIB does not noticeably change as 
column density increases.
We note that the $W_{\rm DIB} = 1.5$~\AA\, value measured by \citet{Geballe_2011_IRdibs}, along
their sightlines with an estimated $A_V \sim 20-30$~mag, falls along the trend
observed in Figure~\ref{fig:ew_vs_extinction}, in the highest $A_V$ regime.
Clearly this DIB carrier is rather robust to conditions along both diffuse and dense sightlines, even where the mean extinction approaches three magnitudes of $V$-band extinction per kiloparsec.

\subsection{Carrier Abundance}
\label{sec:abundance}
Using the relationship to extinction derived above,
we can also compare the estimated carrier abundance (relative to hydrogen) to the estimated abundances of optical DIB carriers. The DIB carrier column density, $N_{\rm car}$, can be expressed as
\begin{equation}
N_{\rm car} = \frac{mc^2}{\pi e^2} \times \frac{W_{\rm DIB}}{\lambda_{\rm DIB}^2} \times \frac{1}{f},
\end{equation}
where $f$ is the transition oscillator strength \citep[e.g.,][]{Spitzer_1978_ISM}. For our feature here (using $W/A_V = 102$~m\AA~mag$^{-1}$; Section~\ref{sec:red}),  the relationship between carrier density and extinction becomes $N_{\rm car}/A_V = 4.58 \times 10^{10}/f $~cm$^{-2}$~mag$^{-1}$. Compared to the mean hydrogen-to-extinction relation of $N_H/A_V = 2 \times 10^{21}$~cm$^{-2}$~mag$^{-1}$ \citep[e.g.,][]{DickeyLockman_90_h1}, we find 
\begin{equation}
N_{\rm car}/N_H \sim 2.3 \times 10^{-11}/f. 
\end{equation}
This is an order of magnitude less abundant than the carriers of the strongest optical DIBs \citep[e.g., $N(5780\normalfont{\AA})/N_H \sim 4 \times 10^{-10}/f$;][]{Tielens_2014_dibs}, though of course both the oscillator strength and the actual number of atoms per carrier molecule are constrained but not actually known for any DIB.

\section{DIB Catalog and Higher-order Moments of the Absorption Field}
\label{sec:feature_properties}

To probe the properties of the diffuse band beyond its zero-th order moment (i.e., its equivalent width), we must restrict our analysis to well-detected individual absorption features. For such systems one can measure the higher-order moments of the absorption line profile to obtain the central wavelength and feature width.

\subsection{Catalog Definition}
\label{sec:catalog}

To select reliably detected DIB features, we start with the list of high quality flux residuals defined above and apply the following additional selection criteria:
\begin{itemize}
\item a feature amplitude $A$ at least three times the local $R_\lambda$ standard deviation. This criterion selects systems in a signal-dominated regime;
\item $\chi^2_\nu$ of the Gaussian DIB fit of 0.95 or smaller.  This choice of criterion (slightly lower than the typical limit of unity) is based on mock absorption feature fits (below, and in Appendix~\ref{sec:dib_fitting_tests}) and ensures that a high-quality fit actually converged; and
\item a feature width $\sigma$ smaller than 8.0~\AA, which excludes a tiny fraction of fits with unphysically large widths (often more than 50\AA), visually confirmed to be caused by, e.g., unusually strong telluric residuals.
\end{itemize}
As a result, we obtain a catalog of DIBs detected in 14,172 sightlines ($\sim$25\% of the ``clean residual'' sample). This catalog increases the number of reliable detections of this DIB feature by roughly three orders of magnitude. 
We have used Monte Carlo simulations with fake absorption features inserted in real spectra
to optimize the above selection criteria and to estimate the reliability and completeness of the catalog. The simulation procedure and reliability assessment are
presented in Appendix~\ref{sec:dib_fitting_tests}, and the completeness estimate is calculated
in Appendix~\ref{sec:completeness}.

\begin{figure*}[ht]
\begin{center}
\includegraphics[trim=0in 0in 0in 0in,clip,width=\textwidth]{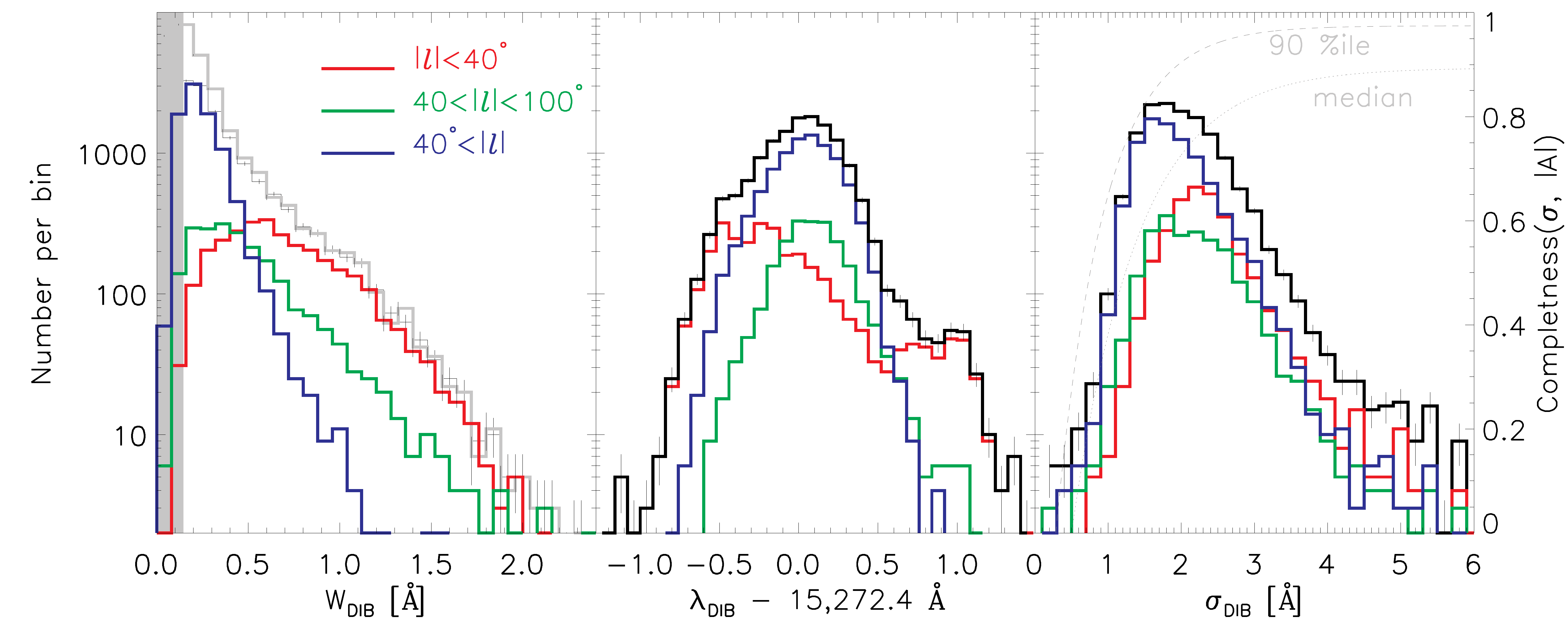}
\caption{
Distributions of the first three moments of the DIB profiles for the entire catalog (black lines) and selected ranges of longitudes (red, green and blue lines):
{\it Left:} Equivalent width $W_{\rm DIB}$. The raw observed distribution is shown in black, and overplotted in gray is the completeness-corrected distribution 
(Appendix~\ref{sec:completeness}). The gray shading marks the range of $W_{\rm DIB}$ where for most observed feature amplitudes, the catalog is less than 20\% complete. 
{\it Middle:} Offset of the measured line center $\lambda$ from the calculated rest wavelength (Section~\ref{sec:rest_wavelength}). The shape of this distribution is driven by the relative velocities of the DIB features, as explored in Section~\ref{sec:kinematics}.
The colors of the lines are the same as in the left panel.
{\it Right:} Profile width $\sigma$.  The colors of the lines are the same as in the left panel.
The right-hand $y$-axis indicates the catalog completeness in $\sigma$ as a function of feature amplitude, $C(\sigma; |A|)$, as traced by the gray lines for two representative amplitudes: dotted for the median $A$ (0.06) and dashed for the 90$^{\rm th}$ percentile value (0.09).
}
\label{fig:param_distributions}
\end{center}
\end{figure*}

The distributions of observed equivalent width $W_{\rm DIB}$, central wavelength $\lambda$ and profile width $\sigma$ for the high-quality catalog features are shown in 
the left, middle, and right panels of Figure~\ref{fig:param_distributions}, respectively.
The distribution of $W_{\rm DIB}$ appears to be well characterized by an exponential distribution. As shown by the colored curves, the characteristic exponential scale depends on Galactic longitude.

\subsection{Determination of the DIB Rest-Frame Wavelength}
\label{sec:rest_wavelength}

Our profile-fitting procedure provides us with an estimate of the observed central wavelength for each system, which can be used to map out the distribution of DIBs in velocity space if the rest-frame wavelength is known. The full distribution of observed peak wavelengths is shown in the middle panel of Figure~\ref{fig:param_distributions}.
The shape of this distribution reflects the variations in radial velocity of the ISM along the line of sight to the background APOGEE stars, which are very dependent on the sightline coordinates and are a strong function of the sampling of different parts of the Galaxy. 

Inferring information on the velocity distribution of the carrier requires a determination of the rest frame wavelength $\lambda_0$ of the DIB transition. Without an identified physical carrier, this rest wavelength must be determined empirically. To do so we can use the following argument: towards the Galactic anti-center at $l=180^\circ$, the expectation value of the ISM line-of-sight velocity is zero,
with a roughly linear trend around the anti-center due to the projection of the ISM's Galactocentric rotation.  We can therefore estimate $\lambda_0$ by looking at the distribution of observed wavelengths around that location.  The presence of significant streaming motions or non-axisymmetric flows in the disk would complicate this assumption, and we check for this possibility by looking for any systematic trend between the DIBs' observed $\lambda$ and (1) the stellar distances 
(as described in Section~\ref{sec:3dmap})
and (2) the line-of-sight extinction. The first is an upper limit on the distance of the DIB carrier distribution, and the second (as shown in Section~\ref{sec:red}) is closely correlated with the integrated absorption $W_{\rm DIB}$ and is a {\it rough} proxy for distance. We do not see a significant trend in either comparison, given the velocity resolution of our measurements. We correct for the measured Galactic $U$ velocity of the Sun by adopting $U=+10.3$~km~s$^{-1}$ (i.e., towards the Galactic Center) from \citet{Bovy_2012_MWrotcurve}.

\begin{figure}[h]
   \includegraphics[angle=90, trim=0.8in 1.5in 1.5in 0in, clip, width=0.5\textwidth]{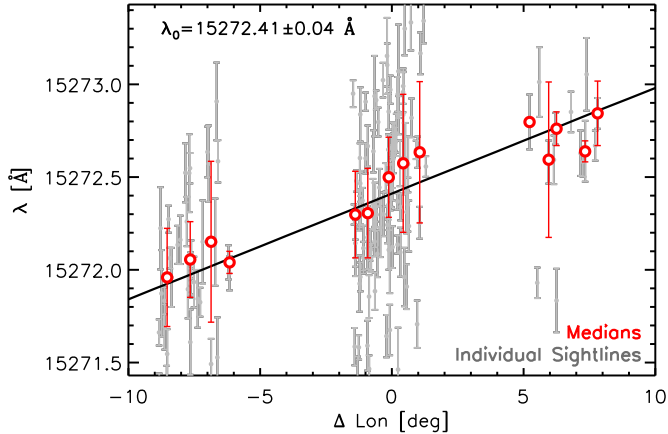} 
\caption{An empirical determination of the DIB's $\lambda_0$, using stars with $170^\circ \le l \le 190^\circ$ (i.e., within 10$^\circ$ of the Galactic anticenter) and $|b| \le 2^\circ$. On the $x$-axis is the distance in longitude from $l=180^\circ$, and the $y$-axis shows the measured central $\lambda$. The gray points are the individual sightline measurements and uncertainties, and the red points indicate the median value and absolute deviation in each $\Delta l = 0.75^\circ$ bin. The solid black line is the best-fitting linear relationship of the individual measurements, whose value at $l = 180^\circ$ is taken as $\lambda_0$.
}
\label{fig:calc_rest_wavelength}
\end{figure}

Figure~\ref{fig:calc_rest_wavelength} shows the distribution of observed wavelengths measured within 10$^\circ$ of the Galactic anti-center in longitude and 2$^\circ$ in latitude. In this region our catalog contains 155 absorber systems 
with background stellar distances $d \le 2$~kpc (to minimize any possible impact from radial flows in the ISM). These points are shown in black, and their median behavior in each $\Delta l = 0.75^\circ$ bin is shown in red. We fit the linear trend of the observed DIB feature wavelengths as a function of angular distance from the anti-center ($\Delta l = l - 180^\circ$):
\begin{equation}
\lambda_{\rm obs} = m\Delta \lambda + \lambda_0,
\end{equation}
where $m$ has units of m\AA~deg$^{-1}$ and is the slope of the wavelength-longitude trend.
We find 
\begin{equation}
\lambda_0 = 15\,272.42 \;\pm\, 0.04~{\rm \AA},
\end{equation}
using a bootstrap resampling of the data to estimate the uncertainties
in the slope and $\lambda_0$.
The slope $m$ across the Galactic anti-center is $57 \pm 8$~m\AA~deg$^{-1}$.
This is within 2$\sigma$ of the values of 
47~m\AA~deg$^{-1}$ from the CO-derived rotation curve of \citet{Clemens_85_COrotcurve},
and 40~m\AA~deg$^{-1}$ from the  stellar-derived curve of \citet{Bovy_2012_MWrotcurve}.
Measuring $m$ and $\lambda_0$ with the anti-center stars on a shorter baseline 
(i.e., with $\Delta l \le 1.5^\circ$)
yields a slope that is much less constrained
and a $\lambda_0$ that is equally well constrained and consistent within the uncertainties.
This well-determined intrinsic wavelength is critical for both identification of the
feature carrier and use of this feature for any kinematical studies of the ISM containing this carrier.

\subsection{Distribution of Carrier Velocities}
\label{sec:kinematics}

\begin{figure}[ht]
   \includegraphics[angle=90, trim=1.2in 1.5in 1.5in 1.5in, clip, width=0.5\textwidth]{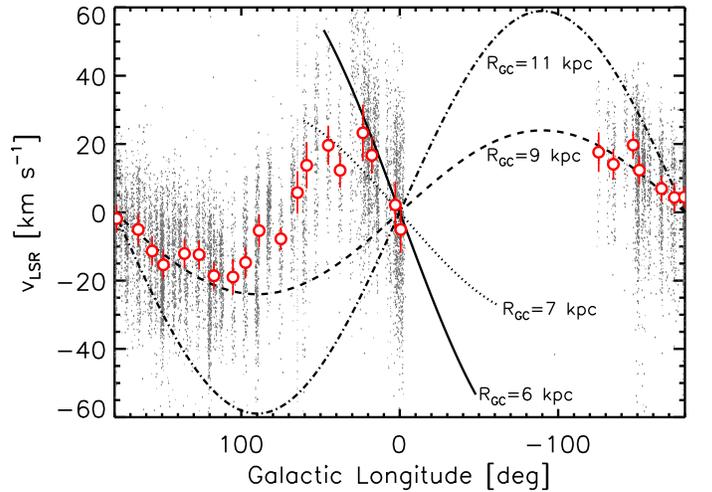}
\caption{
The longitude-velocity ($l$-$v_{\rm LSR}$) diagram of the DIB sample.  Gray points are the measurements along individual sightlines, and the red circles indicate the median and median deviation in each $\Delta l = 10^\circ$ bin. Overplotted are velocity curves from a simple disk rotation model (see text) for Galactocentric circles with $R_{\rm GC} = 6$~kpc (solid), 7~kpc (dotted), 9~kpc (dashed), and 11~kpc (dot-dashed).
}
\label{fig:vlsr}
\end{figure}

Given a reliable measurement of $\lambda_0$,
we can now investigate the distribution of DIB carrier radial velocities relative to the Local Standard of Rest ($v_{\rm LSR}$). Figure~\ref{fig:vlsr} shows the inferred velocities as a function of Galactic longitude. Individual sightlines are shown as gray points, and median values computed in bins of $\Delta l = 10^\circ$ are shown with white circles. Galactic rotation is clearly visible. Also shown are example velocity curves for a simple, axisymmetric disk rotation model \citep[the power-law curve of][]{Bovy_2012_MWrotcurve}, for Galactocentric radii of $R_{\rm GC} = 6$, 7, 9, and 11~kpc. This comparison is not intended to assess the actual kinematic distance distribution of the DIB feature carriers, but rather to demonstrate that the feature velocities and velocity dispersions are consistent with a carrier distribution spanning several kpc of Galactocentric radius.

For sightlines with $|l| \gtrsim 150^\circ$, the observed spread in DIB velocity is relatively small (median deviation of $\sim$8~km~s$^{-1}$), and the velocities themselves are consistent with the model rotation curves for $R_{\rm GC} \sim 9$~kpc (more so than at 11~kpc), suggesting that towards the Galactic anti-center, the DIB carrier density is higher closer to the Sun than in the farther reaches of the disk. In contrast, at inner disk longitudes ($l \lesssim 30^\circ$), the median DIB velocities are more consistent with the model curves at $R_{\rm GC} = 6$~kpc than at 7~kpc, suggesting that the concentration of DIB carrier is higher farther from the Sun, towards the inner Galaxy. These two observations suggest a negative DIB carrier density gradient with $R_{\rm GC}$, a hypothesis we explore in greater detail in Section~\ref{sec:3dmap}. The median deviation in velocities is greater towards the inner disk (1.5$\times$ that of the outer disk), so while the median values themselves show a systematic longitude trend consistent with a range of $R_{\rm GC}$ spanning 6--8~kpc, it is probable that the ISM clouds probed span a much larger range of radii. The departure from the $R_{\rm GC}=9$~kpc curve around $l \sim 90^\circ$ is most likely due to some combination of two factors: First, the distance limit of the stellar sample, which translates into a drastic reduction of APOGEE stars behind $R_{\rm GC} > R_{\rm GC, \sun}$ clouds in the inner Galaxy. At $l = 90^\circ$, points with $R_{\rm GC}=9$~kpc are $\sim$4~kpc from the Sun, increasing to $\sim$10~kpc at $l = 60^\circ$, while the majority of the APOGEE stars in this sample are within $\sim$6~kpc from the Sun.  Second, along sightlines probing inside the solar circle, any weaker features in the outer disk are likely to be dominated by stronger ones in the inner disk.

The middle panel of Figure~\ref{fig:param_map} shows the same $v_{\rm LSR}$ distribution as Figure~\ref{fig:vlsr}, but in a 2D projection in $(l,b)$, where the color of each 0.8~deg$^2$ pixel indicates the median
$v_{\rm LSR}$ value of DIBs in that pixel. Now we can see that the bump in the $\lambda$ distribution seen at longer wavelengths (middle panel of Figure~\ref{fig:param_distributions}) originates from midplane sightlines with $l \sim 30^\circ$. The strong longitudinal variations highlighted in Figure~\ref{fig:vlsr} are clearly visible here, and now with the 2D projection, we can see differences in the longitude behavior at different latitudes. For example, at $l \sim 150^\circ$, the sightlines at $b \sim \pm 4^\circ$ have very similar velocities that are less negative than those in the midplane, but this behavior is not observed at $l \sim -150^\circ$. These kinds of kinematical features will be explored in future work.

\subsection{Distribution of DIB Profile Widths}
\label{sec:fwhm}

As shown in the right panel of Figure~\ref{fig:param_distributions}, the peak of the observed feature width distribution is at $\sigma \sim 1.75$~\AA\,, dominated by sightlines with $|l| > 40^\circ$ (green and blue lines); few sightlines with $|l| < 40^\circ$ (red line) are observed with features this narrow. 
These differences in width may be attributed to an increased distance or velocity dispersion
of ISM clouds along the line of sight, and an increased density of clouds in the inner disk.

The bottom panel of Figure~\ref{fig:param_map} shows the median measured DIB width as a function of $(l,b)$.  On average, the inner Galaxy sightlines contain broader features than the outer Galaxy, but the longitude gradient is not as significant as the one in total $W_{\rm DIB}$ (top of Figure~\ref{fig:param_map}).  Furthermore, we do not see a significant gradient in latitude, even along the minor axis in the bulge, where a $b$-dependent gradient in $W_{\rm DIB}$ is particularly noticeable. This behavior suggests that the patterns in the $W_{\rm DIB}$ distribution are dominated by the feature amplitudes, correlated with carrier density along the line of sight in a smooth manner, rather than discrete clumps.

The right-hand panel of Figure~\ref{fig:param_distributions} also shows the completeness of the DIB catalog feature widths, ($C(\sigma)$; Appendix~\ref{sec:completeness}), for the median and the 90$^{\rm th}$ percentile catalog amplitudes (dotted and dashed gray lines, respectively). The completeness for deeper features (larger $|A|$) is higher at all widths, as expected, but we note that even this one decreases rapidly for $\sigma \lesssim 1.75$~\AA --- i.e., at the peak of the catalog distribution. Thus we cannot conclude that our narrowest features represent the intrinsic width of this DIB.

\section{Towards Mapping the Carrier's 3D Distribution}
\label{sec:3dmap}

We now consider the mapping of the DIB carrier throughout the Galaxy. A characterization of its 3D distribution can provide us with useful insight on the DIB carrier host environment --- e.g., How far above the disk does the carrier extend?  How well-mixed it is in the midplane? ---
as well as on the large-scale distribution of metals in the Galaxy. To approach these questions, we use additional sightline information --- namely, the background star distances from \citet[][and in preparation]{Hayden_2013_chemicalcartography}, who adopt a Bayesian approach using the stellar photometry, spectroscopically-derived stellar parameters, theoretical stellar models, and Galactic stellar density priors. 

The spectroscopic stellar parameters used in the distance calculations come from the 
best-fitting synthetic spectra produced by the ASPCAP pipeline and used in 
Section~\ref{sec:residuals}.  In addition to templates for the stellar contribution to the total observed spectra, these also provide us with estimates of physical parameters for the stars: effective temperature ($T_{\rm eff}$); surface gravity ($\log{g}$); metallicity ([Fe/H]); and relative abundances of carbon, nitrogen, and the $\alpha$ elements ([C/M], [N/M], [$\alpha$/M]). \citet{Meszaros_2013_aspcapcalib} discuss the calibration of these parameters using cluster and field stars with high-quality parameters in the literature derived from high-resolution optical spectroscopy and asteroseismology; after small empirical adjustments, they conclude the ASPCAP values (within known bounds) are reliable to within approximately $\pm$150~K in $T_{\rm eff}$, $\pm$0.2 dex in $\log{g}$, and $\pm$0.1 dex in [Fe/H]. 
We impose requirements on these parameters to ensure that the derived distances are reliable.

For ASPCAP to successfully characterize a star, the spectral S/N must be high and the stellar parameters must lie within the ranges spanned by the synthetic model grid and empirically calibrated as described above. These factors impose requirements of $({\rm S/N)}_{F_\lambda} \ge 70.0$ per spectrum pixel, $3500 \le T_{\rm eff} \le 5500$~K, and $\log{g} \le 3.8$. We also require that the ASPCAP ``WARN'' and ``BAD'' flags for the $T_{\rm eff}$, $\log{g}$, [Fe/H], and $\chi^2$ all be set to 0, and the distance uncertainties be $\le$50\%. These requirements are added to those on $F_\lambda$ and $R_\lambda$ described in Section~\ref{sec:aspcap}, but as in Section~\ref{sec:mean_properties}, we do not place any explicit requirements on the DIB profile or fit itself.
For visual clarity, in this section's plots we show only sightlines with $|b| \le 25^\circ$, though all sightlines meeting the above criteria are used in the calculations.

\begin{figure*}[!ht]
\includegraphics[angle=90, trim=1.3in 4.5in 1.4in 1.0in, clip, width=0.5\textwidth]{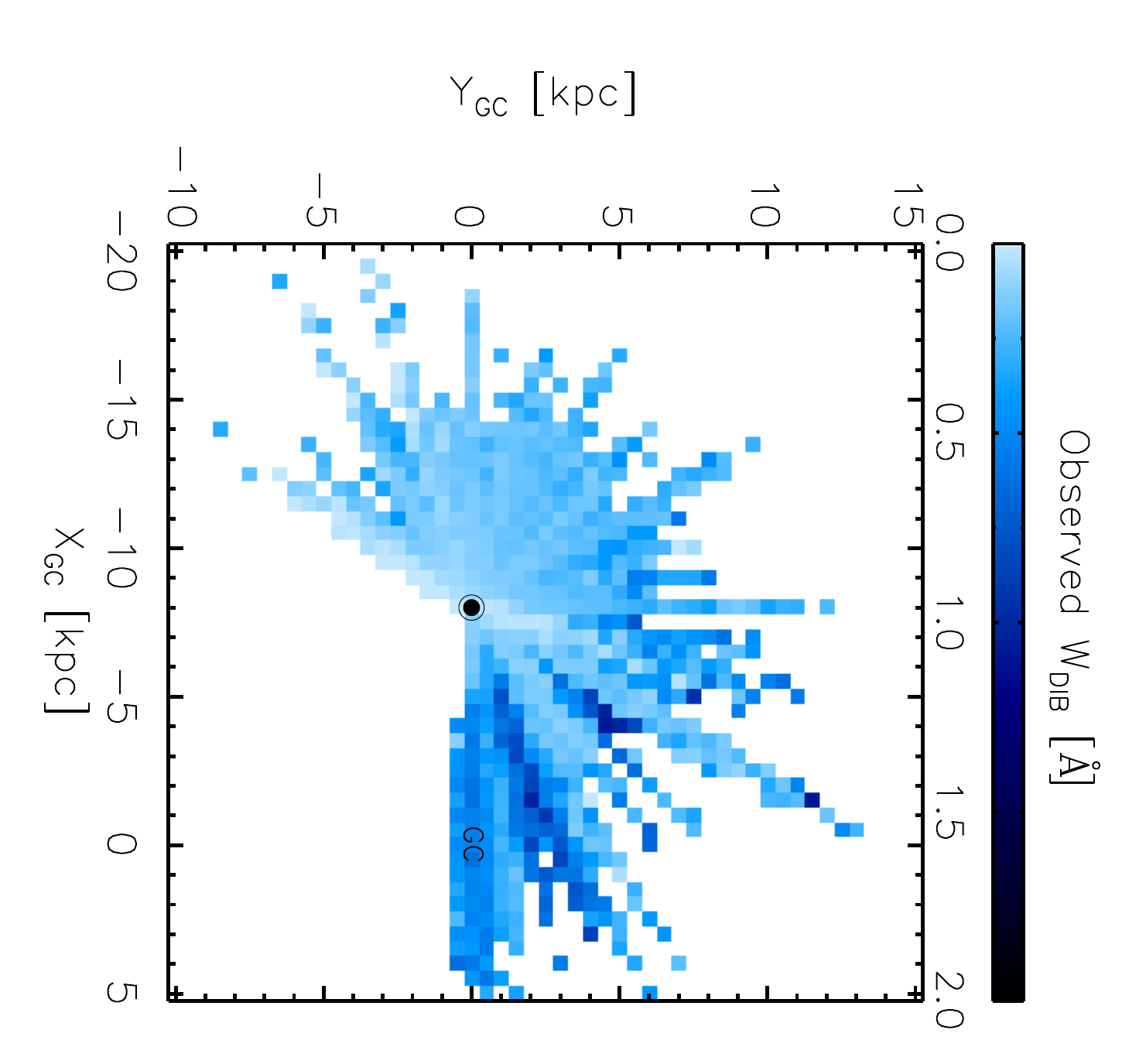}
\includegraphics[angle=90, trim=1.3in 4.5in 1.4in 1.0in, clip, width=0.5\textwidth]{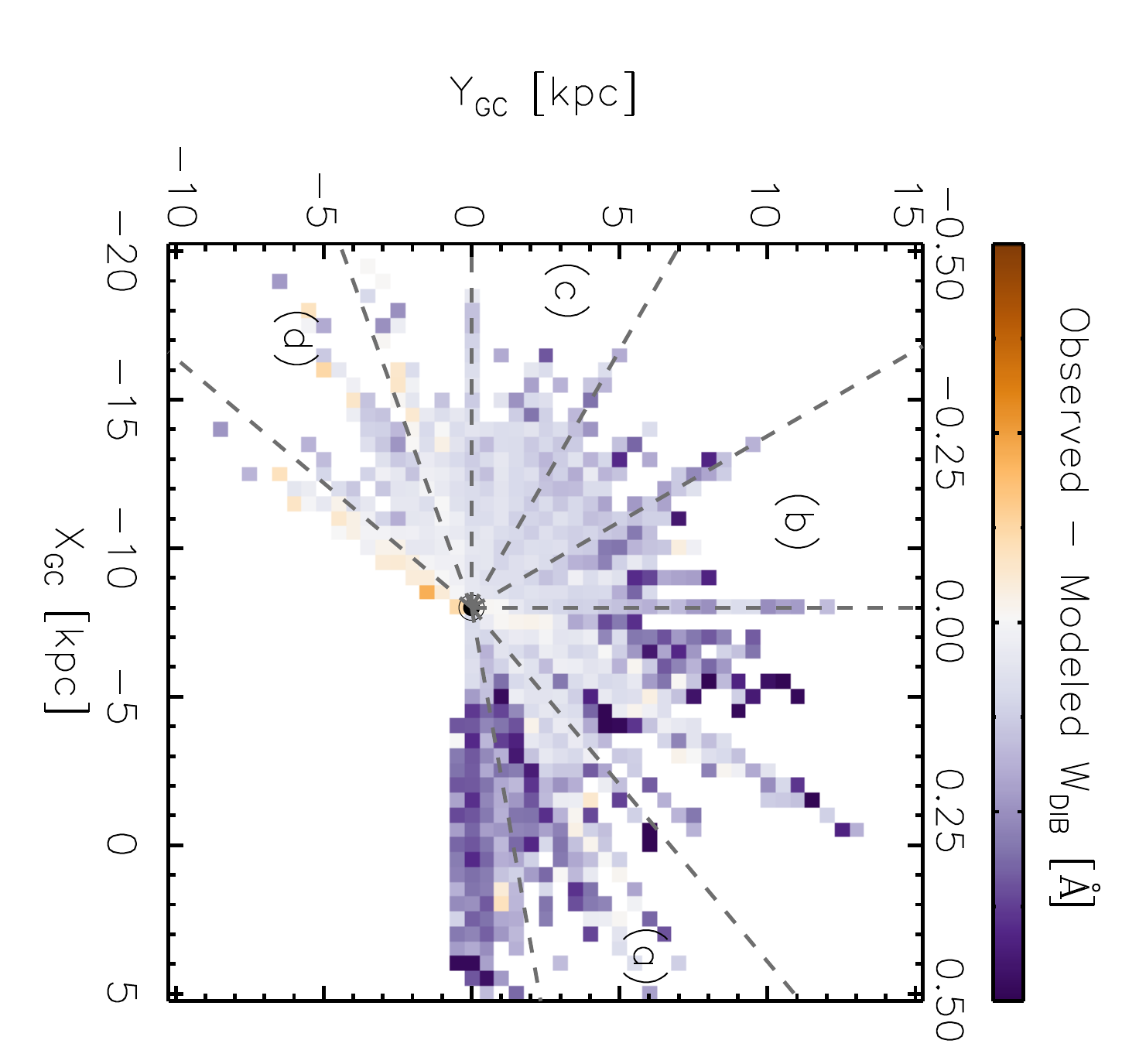} 
\caption{
{\it Left}: Pole-down view of the Galactic disk, showing the median observed $W_{\rm DIB}$ of sightlines terminating in each 0.25~kpc$^{2}$ pixel.
{\it Right}: Difference between the median observed $W_{\rm DIB}$ distribution in the Galactic disk and that predicted by the best-fitting smooth exponential disk model,
for stellar sightlines in each pixel.
The lettered zones refer to the panels in Figure~\ref{fig:xymap_rzdiff}.
}
\label{fig:xymap_obs_diff}
\end{figure*}

The distances for the resulting sample of 49,044 stars span 1--15~kpc, with a mean of $\sim$3.5~kpc and typical uncertainties between 20-30\%. The spatial distribution of these stars is shown in Figure~\ref{fig:xymap_nstars}. In the left panel of Figure~\ref{fig:xymap_obs_diff}, 
we show the distribution of the DIB absorption field's integrated equivalent widths, 
as described in Section~\ref{sec:absorption_field};
the color scale indicates the median $W_{\rm DIB}$ value measured from the APOGEE spectra of the stars in each 0.25~kpc$^{-2}$ pixel. 
We emphasize that the distances shown in this map are those of the background stars, 
which are an upper limit on the typical distance (or distance range) of each sightline's DIB carrier.
Other approaches to mapping the DIB density directly (e.g., using kinematical distances or differential line-of-sight absorption intensity) 
require additional assumptions and will be explored in the future.
But even in this projection, disk-scale gradients from the inner to outer Galaxy are visible.
Much of the apparent variance here is due to the uneven sampling of the Milky Way by APOGEE
sightlines --- for example, the lighter colored (lower $W_{\rm DIB}$) stripe extending from
the Sun to $(X_{\rm GC},Y_{\rm GC}) \sim (0,12)$ 
is due to a significantly larger fraction of higher-latitude
stars at those longitudes, pulling the median $W_{\rm DIB}$ in those pixels to lower values.

To characterize the spatial distribution of the DIB carrier, 
we use a toy model comprising a simple exponential disk, parameterized\footnote{Adding a fourth free parameter $Z_C$, the vertical offset between the DIB disk midplane and the Sun's position, does not change the resulting fit parameters, so for the sake of simplicity we fix it to $Z_C = -25$~pc 
\citep[as measured in both stars and gas; e.g.,][]{Sanders_1984_CO-H2distribution,Juric_2008_MWstellardensity}.} by 
a scale length $h_R$, scale height $h_Z$ and normalization $\kappa_0$. For every star at heliocentric distance $d$, the observed equivalent width $W_{\rm DIB}$ is the integration of absorption along the {\it total} line of sight to the star:
\begin{equation}
W_{\rm model}(R,Z) = \kappa_0 \int_0^d e^{-R/h_R} \, e^{-|Z|/h_Z} \, {\rm d}s.
\end{equation}
We find the distribution of the DIB carrier to be best represented by the following parameters:
\begin{eqnarray}
{\rm scale\;height:}~h_Z &=& 108\pm 8~{\rm pc}, \nonumber\\
{\rm scale\;length:}~h_R &=& 4.9\pm 0.2~{\rm kpc}, 
\end{eqnarray}
and a normalization $\kappa_0 = 0.55\pm 0.04$~\AA~kpc$^{-1}$ at the Galactic Center. 
The uncertainties on the parameters are estimated from a bootstrap resampling (with replacement)
of the data.
To evaluate the uncertainty induced by uncertainties in the stellar distances, which are not formally included in the fits, we performed an additional series of 100 model fits, varying each time the stellar distances using a normal distribution representing their individual uncertainties.  We find the variation in the resulting series of fit parameters to be much less than the bootstrap-derived parameter uncertainties, so we use the latter alone for our result.

To place these structural parameters in context, we note that dense molecular gas as traced by $^{12}$CO has a scale height of $\sim$50--70~pc \citep[increasing with Galactocentric radius;][]{Sanders_1984_CO-H2distribution}; \citet{Langer_2014_CIIscaleheight} recently measured the mean scale height of [\ion{C}{2}] (tracing both atomic and molecular clouds) to be 73~pc. Constraining the scale height of the atomic \ion{H}{1} layer is complicated by its structure and asymmetries  (including the prominent warp and flare), but typical values in the range spanned by the APOGEE sample are around $\sim$200~pc \citep[approximated from the components measured by][]{DickeyLockman_90_h1}, 
larger than measurements of the dust layer \citep[e.g., $\sim$125~pc;][]{Marshall_06_3Dextinction}. We find the DIB carrier, therefore, to have a vertical profile between that of atomic hydrogen (and dust) and material residing (at least in part) in molecular clouds.

The scale length is comparable to that adopted for the ISM in the Besan\c{c}on Galaxy Model \citep[$\sim$4.5~kpc;][]{Robin_03_besanconmodel}. It is rather larger than typically measured for the stellar thin and thick disks \citep[$\sim$1.8--3.6~kpc; e.g.,][]{Juric_2008_MWstellardensity,Cheng_2012_thickdiskscalelength}, though \citet{Bovy_2012_diskpops} report a scale length of 4.3~kpc for their most $\alpha$-poor dwarf sample.

\subsection{Revealing Galactic Substructure}

\begin{figure*}[!htbp]
\includegraphics[trim=4.2in 1in 1.0in 0.8in, clip, angle=90, width=\textwidth]{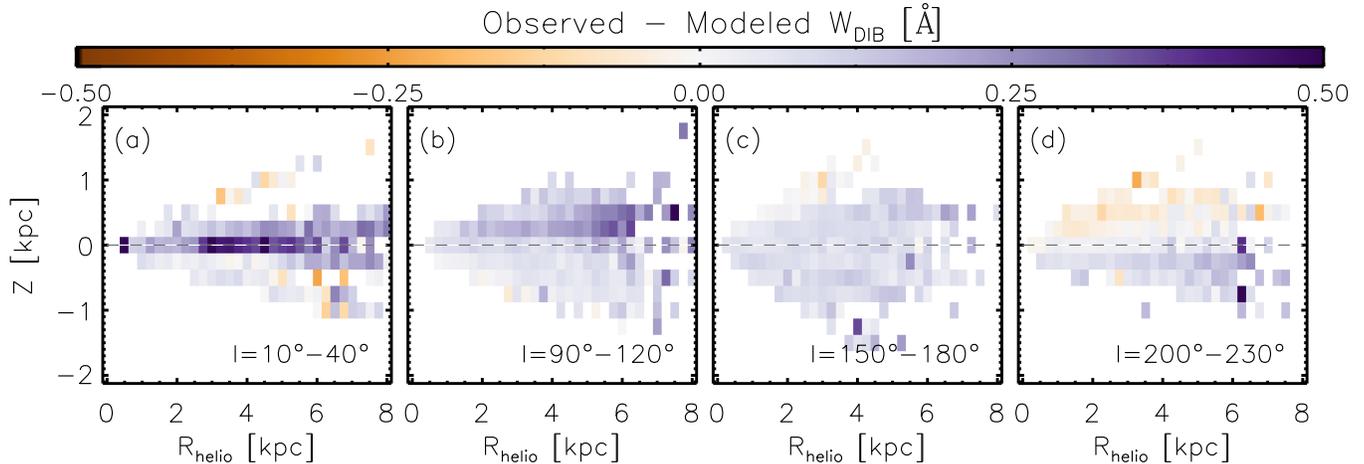}
\caption{
Difference between the median observed $W_{\rm DIB}$ distribution in the Galactic disk and that predicted by the best-fitting smooth exponential disk model, as a function of projected midplane distance $R_{\rm helio}$ and vertical height above the plane $Z$ of the stars.  Each panel corresponds to the range of longitude indicated.
}
\label{fig:xymap_rzdiff}
\end{figure*}

In addition to estimating the global density distribution of the DIB carrier, we also look for localized departures from a smooth and symmetric model --- due, for example, to the presence of the inner Galactic bar/bulge structure and the Galactic disk warp and flare.  We compute the difference between the observed distribution of DIB absorption strength and the best-fit model derived above:
\begin{equation}
\Delta W= W_{\rm obs} - W_{\rm model}\;.
\end{equation}
This quantity is shown in the right panel of Figure~\ref{fig:xymap_obs_diff} 
in the same projection of the disk as the left panel; 
the letters refer to the individual parallel-to-the-disk views
in Figure~\ref{fig:xymap_rzdiff} (described below).
Note that the total scale of $\Delta W$ shown spans about half of the scale
used in the left panel of Figure~\ref{fig:xymap_obs_diff}'s $W_{\rm DIB}$ map.

The most prominent deviations occur towards the Galactic Center ($X_{\rm GC},Y_{\rm GC} = 0,0$), starting about 3~kpc from the Sun, and along the line of sight at $l \sim 20-30^\circ$ between 3--5~kpc from the Sun. This extended overdensity may be due to an inability of the {\it global} model parameters to describe the local density distribution of the Galactic bulge. To explore this possibility, we fit the model disk again using only sightlines with $|l| \le 30^\circ$, which yields a similar scaleheight as the total fit, but a significantly longer scalelength of 9.63~kpc and a  lower normalization (0.4~\AA~kpc$^{-1}$). Repeating the fit with similarly-sized $|l|$ ranges centered at $l = 85^\circ$ and 180$^\circ$ shows that the DIB distribution parameters are largely consistent throughout the Milky Way beyond $l \sim 30^\circ$. This consistency, and the deviation observed within $|l| \le 30^\circ$, suggests that a more complex distribution function or a  multi-component model is required to trace fully the DIB carrier component of the ISM. 

The stronger DIB absorption observed between $20^\circ \lesssim l \lesssim 30^\circ$ does not disappear even when the longer scalelength predicted by the inner disk absorption is taken into account. 
The panels in Figure~\ref{fig:xymap_rzdiff} show the distribution of $\Delta W$  (from the model fitted to all sightlines; only those with $|b| \le 30^\circ$ are shown) as a function of projected stellar midplane distance ($R_{\rm helio}$) and height above the midplane ($Z$) for a few longitude ranges; the letters in these panels match the lettered longitude ranges
in the right panel of Figure~\ref{fig:xymap_obs_diff}.
Figure~\ref{fig:xymap_rzdiff}a focuses on the inner $l$ sightlines.  We can see that this strong $\Delta W$ is highly concentrated in the midplane, and between 3--5~kpc from the Sun. This overdensity is spatially coincident with a number of known Milky Way features: the end of the central ``long bar'' \citep{Benjamin_05_glimpse,Zasowski_2012_innerMW}, the bar's intersection with the inner Scutum-Centaurus and Norma spiral arms, and the extensive Scutum Complex of star-forming regions  thought to be at least partially driven by the dynamics of this intersection \citep[e.g.,][]{L-C_1999_scutumSFR}. The chemistry and radiation field of this region apparently also support a comfortable environment for the 1.527~$\mu$m DIB carrier; since these chemical and radiative properties are constrained by numerous observations of stellar and ISM components of the region, this 3D mapping of the DIB carrier can be used to place stronger constraints on the properties (composition, ionization, durability) of the DIB carrier itself.
We note that the $\Delta W \sim 0.5$~\AA~pixel at $(R, Z) = (0.5, 0)$ is caused by a small number of short sightlines that dominate that particular location, 
where the total number of sightlines is small; in the comparable location in the right panel of Figure~\ref{fig:xymap_obs_diff} ($[X_{\rm GC},Y_{\rm GC}] \sim [-7.5, 0.5]$),
these sightlines are binned with others of lower $\Delta W$ that reduce these outliers' influence on the measured median.

Also discernible in the comparison of the observed $W_{\rm DIB}$ spatial distribution to that predicted by the best-fitting disk model is the  Galactic disk's warp.  The line of nodes of this structure is near $l \sim 180^\circ$, with apparent maximum departure from $b=0^\circ$ of about +2$^\circ$ at $l \sim 90^\circ$ and about --2$^\circ$ at $\sim$270$^\circ$.  Figures~\ref{fig:xymap_rzdiff}b and \ref{fig:xymap_rzdiff}d show the  distribution of $\Delta W$ for these two longitude ranges (in the latter case, as close as the Northern hemisphere APOGEE data can reach).  The systematic over-absorption for $b>0^\circ$ at $l \sim 90^\circ$  and for $b<0^\circ$ at $l \sim 230^\circ$ (compared to the unwarped disk model) is consistent with the DIB carrier being distributed in a warped disk as well. A quantitative characterization of the warp parameters and comparison to the stellar and \ion{H}{1} distributions are left for future work.

Finally, in Figure~\ref{fig:xymap_rzdiff}c, we highlight the $R-Z$ distribution of $\Delta W$ approaching the disk anti-center. Because this is near the warp's line of nodes, we do not see strong asymmetric (in latitude) departures from the smooth model, but we note that $\Delta W > 0$~\AA\, differences are apparent both above and below the midplane, with $\Delta W \sim 0\,$\AA\ right around $Z \sim 0$~kpc. This behavior is consistent with a {\it flared} DIB distribution, since a smooth model with a constant scaleheight will increasingly under-predict the $|Z| \neq 0$~kpc sightline absorption (and over-predict the midplane $Z=0$~kpc absorption). The fact that the DIB carrier follows this flare,  extending up to $\geq 1-1.5$~kpc from the midplane in the outer reaches of the Galactic disk,  means that it is well-mixed in the diffuse ISM (consistent with its correlation with dust; Section~\ref{sec:red}); combined with the carrier's strength in the very different environments of the inner disk, its presence also in the outer disk also means that the carrier is prevalent in a very wide range of radiation conditions.

\section{Conclusions}

We have presented an exploration of the distribution and properties of 
the interstellar medium of the Milky Way as traced by diffuse interstellar bands,
absorption features that are produced by molecules likely containing large numbers of carbon and other metal atoms.
To do so we have analyzed near-infrared stellar spectra from the SDSS-III's APOGEE survey to detect the absorption band at $\lambda \sim 1.527 \,\mu{\rm m}$, the strongest DIB in the $H$-band. The APOGEE survey probes stars in a wide range of Galactic environments, from the halo to the heavily-obscured bulge and inner disk, spanning distances up to 15 kpc from the Sun and probing up to 30 magnitudes of visual extinction. The sample comprises mostly K and early M giants ($T_{\rm eff}\sim 3500-5000$~K), whose spectra are dominated by stellar atmospheric absorption lines. Reliable removal of these lines is crucial for identifying interstellar absorption features. We have made use of stellar continuum and absorption line estimates from ASPCAP, 
together with a set of additional masks and median filters, 
to produce residuals free of stellar or terrestrial contamination
in which to measure interstellar absorption.

We have characterized the DIB absorption field in both the signal- and noise-dominated regimes from the analysis of about 58,000 residuals. Our results can be summarized as follows:
\begin{itemize} \itemsep -1pt
\item We have created a set of maps of the 1.527~$\mu$m DIB absorption towards the Galactic plane, showing its mean spatial distribution of strength, line-of-sight velocity, and profile width, each of which reveals certain aspects of the structure of the Galaxy.
\item We find the DIB strength to be linearly correlated with dust extinction over three orders of magnitude in column density.
A power-law fit has an index of $1.01 \pm 0.01$ and
an amplitude of $W_{\rm DIB}$/$A_V = 102 \pm 1$~m\AA~mag$^{-1}$.  
The linearity of this relationship over such a wide range 
of reddening, coupled with the strength of the feature relative to the reddening, 
establishes this DIB as a powerful, independent tracer of interstellar dust.
\item We have created a catalog of robustly detected DIB features, for which the central wavelength and line width can be measured, in about 14,000 sightlines. The longitude dependence of the DIB velocity distribution reveals the kinematical structure of the rotating Galactic disk.
\item Using the expected symmetry of the ISM velocity distribution around the Galactic anti-center, we have empirically estimated the rest-frame wavelength of the DIB. We find $\lambda_0 = 15\,272.42 \pm 0.04$~\AA.
\item With information on the stellar distances, we have mapped out the 
integrated line-of-sight DIB absorption and showed that the DIB carrier distribution can be characterized by an exponential disk model with a scale height of about 100 pc 
(comparable to other ISM tracers; e.g., \ion{H}{1}, CO, [\ion{C}{2}], dust) and a scale length of about 5 kpc (more extended than the stellar disks). Finally, we show that the integrated DIB absorption distribution also traces large-scale Galactic structures, including the Galactic long bar and the warp of the outer disk.
\end{itemize}

This work highlights the potential of near-IR DIBs to be powerful tracers of metals in the ISM, able to probe a broad range of Galactic environment and provide us with information on the three dimensional distribution of the gaseous component of the Milky Way, at distances reaching from the central bulge to the edge of the outer disk.

\begin{acknowledgments}

GZ is supported by an NSF Astronomy \& Astrophysics Postdoctoral Fellowship under Award No.\ AST-1203017. This work was also supported by NSF Grant AST-1109665. 
We thank D.~Hogg, J.~Bovy, M.~Schultheis, and D.~York for useful discussion and feedback,
and the anonymous referee for comments that improved the clarity of the paper.

Funding for SDSS-III has been provided by the Alfred P. Sloan Foundation, the Participating Institutions, the National Science Foundation, and the U.S. Department of Energy Office of Science. The SDSS-III web site is \url{http://www.sdss3.org/}.

SDSS-III is managed by the Astrophysical Research Consortium for the Participating Institutions of the SDSS-III Collaboration including the University of Arizona, the Brazilian Participation Group, Brookhaven National Laboratory, University of Cambridge, Carnegie Mellon University, University of Florida, the French Participation Group, the German Participation Group, Harvard University, the Instituto de Astrofisica de Canarias, the Michigan State/Notre Dame/JINA Participation Group, Johns Hopkins University, Lawrence Berkeley National Laboratory, Max Planck Institute for Astrophysics, Max Planck Institute for Extraterrestrial Physics, New Mexico State University, New York University, Ohio State University, Pennsylvania State University, University of Portsmouth, Princeton University, the Spanish Participation Group, University of Tokyo, University of Utah, Vanderbilt University, University of Virginia, University of Washington, and Yale University. 

\end{acknowledgments}

\bibliographystyle{aa}
\bibliography{../../../reflib.bib}

\begin{appendix}

\section{Robustness of DIB Profile Fits} 
\label{sec:dib_fitting_tests} 

To test the reliability of our DIB profile fitting algorithm,  we performed two additional analyses using simulated DIB features --- that is, we added Gaussian absorption profiles to the normalized $F_\lambda$ and explored how well the fitting algorithm  could recover the input parameters ($A_0$, $\lambda_0$, and $\sigma_0$) from the resulting $R_\lambda$. The simulated profile for each sightline used parameters selected  within a range extending beyond that observed in the APOGEE residuals --- $0 \le |A_0| \le 0.1$ and $0.1 \le \sigma_0 \le 9$~\AA\, --- with distributions skewed towards weaker $A_0$ and smaller $\sigma_0$, where the majority of the observed features lie.

In the first case, we drew a random subset of 10,000 residual spectra and added these mock DIB profiles within $1.5475 \le \lambda_0 \le 1.5495$~$\mu$m, a region without any noticeable residual features  but with similar local variance and sky emission line density to the region surrounding the 1.527~$\mu$m DIB. In the second case, we added simulated profiles around the expected position of the real feature ($1.5262 \le \lambda_0 \le 1.5282$~$\mu$m) in $\sim$1300 high-latitude stars ($|b| \ge 60^\circ$) in the sample  with ``clean'' residuals and without any measurable excess absorption in this wavelength range ($|A| \le  \sigma(R_\lambda)/2$). This case tests the fitting procedure on simulated profiles with the same nearby stellar and sky residuals on which the actual DIB features are superimposed. We ran this latter case 100 times, with a different mock DIB profile feature added to each star's normalized $F_\lambda$ spectrum each time, for a total of nearly 130,000 mock features added to this high-$b$ stellar sample.

\begin{figure*}[ht]
\begin{center}
  \includegraphics[trim=2.5in 0.5in 2in 0.5in, clip, angle=90, width=\textwidth]{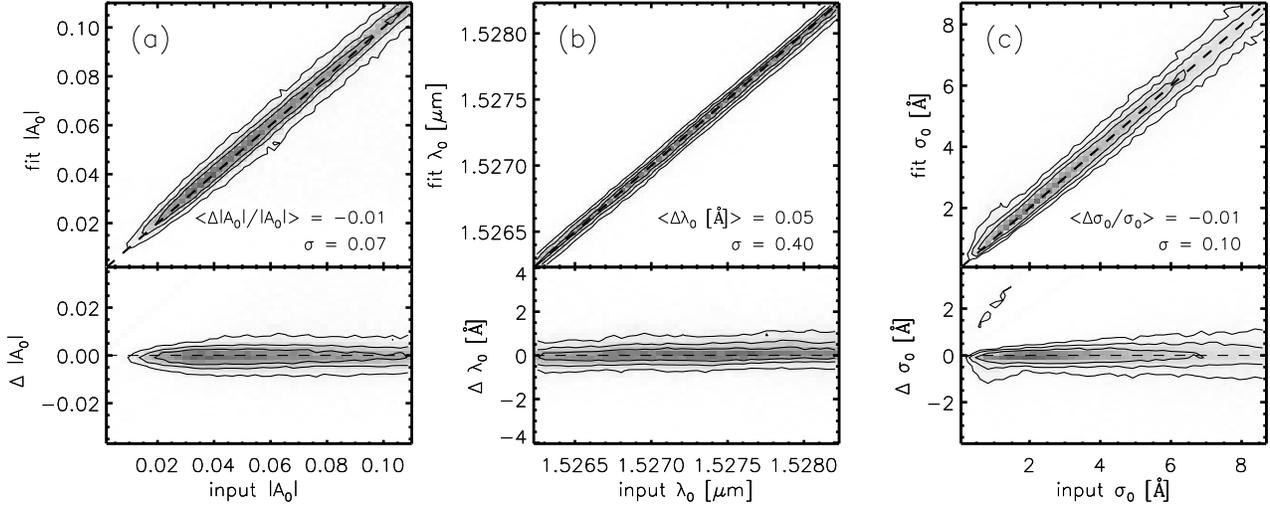} 
\caption{Comparison of the input and best-fit parameters of the mock DIB profiles: {\it (a)} amplitude $|A_0|$, {\it (b)} central wavelength $\lambda_0$, and {\it (c)} profile width $\sigma_0$. The top panels show the direct comparison and the 1:1 relation (dashed line), and the bottom panels show the difference ($\Delta = {\rm input - fit}$) as a function of input parameter. The contours enclose 50\%, 75\%, and 90\% of the sample. Also shown are median fractional differences for $A_0$ and $\sigma_0$, the median absolute difference for $\lambda_0$,
and the dispersion in each case.
}
\label{fig:mockdibfits}
\end{center}
\end{figure*}

We compared the difference between the input and fitted parameters to a variety of stellar and spectral properties. Both tests yielded very similar results, and we show the outcome from the second case in Figure~\ref{fig:mockdibfits}, for those spectra in which a fit actually converged ($\sim$47,000 spectra). We find no significant correlation between the fitting accuracy and the stellar parameters (where reliable stellar parameters are available to be compared), which indicates that we are not biased in measuring DIBs towards certain kinds of stars, provided the synthetic spectral model is a good match to the spectrum, at least locally. In a comparison of the input/fitted parameters themselves, the only systematic trend is  for features with very narrow input widths ($\sigma_0 \lesssim 1.5$~\AA) to have sightly underestimated amplitudes ($\Delta A/A_0 \sim 20\%$) and overestimated fitted widths ($\Delta \sigma/\sigma_0 \sim 5\%$). This is most likely due to the similarity between these narrow lines and residual {\it stellar} features,  and the fitting routine's attempt to fit the apparent complex of residual lines as a single broad, shallow feature.
Coincidentally, the total measured equivalent widths in these cases is very close to the actual
input equivalent width;
nevertheless, in the observed sample, if such narrow intrinsic widths exist in the spectra, the majority of them will be rejected from
the catalog of detected features because of their low amplitudes.

\section{Completeness of the High S/N DIB Catalog} 
\label{sec:completeness} 

\begin{figure}[!htbp]
   \includegraphics[trim=1.0in 1.5in 1in 2.9in, clip, angle=90, width=0.5\textwidth]{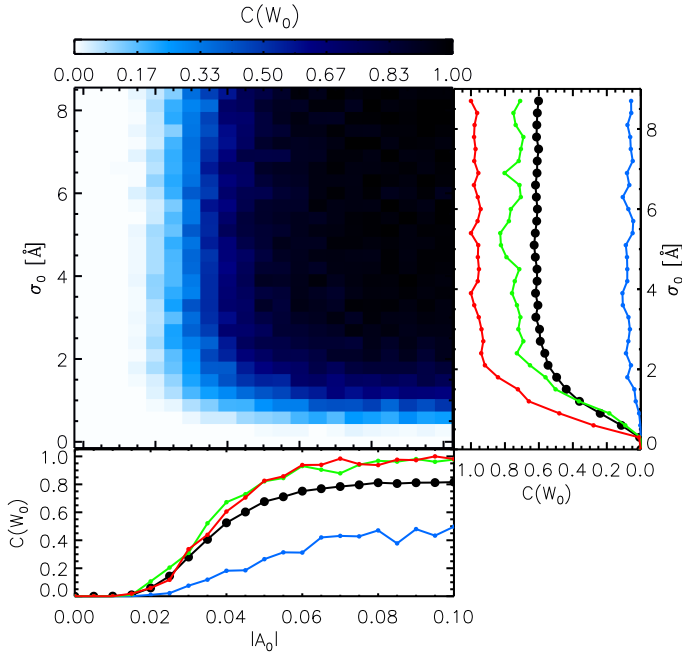} 
\caption{
$W_{\rm DIB}$ completeness estimate over the $A_0$ and $\sigma_0$ ranges tested using mock DIB features. The blue shaded panel indicates the fraction of of well-detected features in the input $A_0-\sigma_0$ field (see text). {\it Bottom:} Completeness as a function of $|A_0|$, averaged over the range of input $\sigma_0$ (black) and for  three sample $\sigma_0$ values (blue: 0.9~\AA, green: 3.6~\AA, red: 7.5~\AA). {\it Right:} Completeness as a function of $\sigma_0$, averaged over the range of input $|A_0|$ (black) and for three sample $|A_0|$ values (blue: 0.02, green: 0.05, red: 0.09).
}
\label{fig:completeness}
\end{figure}

We calculate the fraction of mock profiles (Appendix~\ref{sec:dib_fitting_tests}) whose fits not only meet the catalog feature requirements given in Section~\ref{sec:catalog} but also recover both amplitude and width to within 50\% of the feature input values.   This fraction, $C(W_0)$, is shown in Figure~\ref{fig:completeness} in shaded bluescale, as a function of input $|A_0|$ and $\sigma_0$.  The panel along the bottom shows $C(W_0)$ as a function of $|A_0|$, averaged over the full range of $\sigma_0$ (black points) and for some example $\sigma_0$ values. The panel along the right side shows comparable relationships for $C(W_0)$ as a function of $|A_0|$. In the left panel of Figure~\ref{fig:param_distributions}, we use the two dimensional, $A$- and $\sigma$-dependent $C(W)$ estimate to correct the distribution for completeness based on each feature's measured $A$ and $\sigma$.

\end{appendix}

\end{document}